\documentclass[twocolumn]{aastex61}
\pdfoutput=1 
\usepackage{amsmath,amstext}
\usepackage[T1]{fontenc}
\usepackage{apjfonts} 
\usepackage[figure,figure*]{hypcap}
\usepackage{gensymb}
\usepackage{ amssymb }
\usepackage{float}

\shorttitle{The nuclear star cluster of Sgr dSph}
\shortauthors{M. Alfaro-Cuello et al.}

\begin{document}

\title{A deep view into the nucleus of the Sagittarius Dwarf Spheroidal Galaxy with MUSE.\\ 
II. Kinematic characterization of the stellar populations.}

\author{M. Alfaro-Cuello}
\affiliation{Max-Planck-Institut f\"ur Astronomie, K\"onigstuhl 17, 69117 Heidelberg, Germany.}

\author{N. Kacharov}
\affiliation{Max-Planck-Institut f\"ur Astronomie, K\"onigstuhl 17, 69117 Heidelberg, Germany.}

\author{N. Neumayer}
\affiliation{Max-Planck-Institut f\"ur Astronomie, K\"onigstuhl 17, 69117 Heidelberg, Germany.}

\author{P. Bianchini}
\affiliation{Observatoire astronomique de Strasbourg, CNRS, UMR 7550, F-67000 Strasbourg, France.}

\author{A. Mastrobuono-Battisti}
\affiliation{Max-Planck-Institut f\"ur Astronomie, K\"onigstuhl 17, 69117 Heidelberg, Germany.}

\author{N. L\"utzgendorf}
\affiliation{European Space Agency, c/o STScI, 3700 San Martin Drive, Baltimore, MD 21218, USA.} 

\author{A.C. Seth}
\affiliation{Department of Physics and Astronomy, University of Utah, Salt Lake City, UT 84112, USA.}

\author{T. B\"oker}
\affiliation{European Space Agency, c/o STScI, 3700 San Martin Drive, Baltimore, MD 21218, USA.} 

\author{S. Kamann}
\affiliation{Astrophysics Research Institute, Liverpool John Moores University, 146 Brownlow Hill, Liverpool L3 5RF, United Kingdom.}

\author{R. Leaman}
\affiliation{Max-Planck-Institut f\"ur Astronomie, K\"onigstuhl 17, 69117 Heidelberg, Germany.}

\author{L. L. Watkins}
\affiliation{European Southern Observatory, Karl-Schwarzschild-Str. 2, 85748 Garching, Germany.}
\affiliation{Department of Astrophysics, University of Vienna, T\"urkenschanzstrasse 17, 1180 Wien, Austria.}

\author{G. van de Ven}
\affiliation{Department of Astrophysics, University of Vienna, T\"urkenschanzstrasse 17, 1180 Wien, Austria.}

\begin{abstract}

The Sagittarius dwarf spheroidal galaxy (Sgr dSph) is in an advanced stage of disruption but still hosts its nuclear star cluster (NSC), M54, at its center. 
In this paper, we present a detailed kinematic characterization of the three stellar populations present in M54: young metal-rich (YMR); intermediate-age metal-rich (IMR); and old metal-poor (OMP), based on the spectra of $\sim6500$ individual M54 member stars extracted from a large MUSE/VLT dataset.
We find that the OMP population is slightly flattened with a low amount of rotation ($\sim0.8$~km~s$^{-1}$) and with a velocity dispersion that follows a Plummer profile. The YMR population displays a high amount of rotation ($\sim5$~km~s$^{-1}$) and a high degree of flattening, with a lower and flat velocity dispersion profile. The IMR population shows a high but flat velocity dispersion profile, with some degree of rotation ($\sim2$~km~s$^{-1}$).
We complement our MUSE data with information from \textit{Gaia DR2} and confirm that the stars from the OMP and YMR populations are comoving in 3D space, suggesting that they are dynamically bound. While dynamical evolutionary effects (e.g. energy equipartition) are able to explain the differences in velocity dispersion between the stellar populations, the strong differences in rotation indicate different formation paths for the populations, as supported by an $N$-body simulation tailored to emulate the YMR-OMP system. This study provides additional evidence for the M54 formation scenario proposed in our previous work, where this NSC formed via GC accretion (OMP) and in situ formation from gas accretion in a rotationally supported disc (YMR).

\end{abstract}


\section{Introduction}

Nuclear star clusters (NSCs) are the densest stellar systems known in the Universe with mass densities of $\sim10^{6-7}\,M{_\odot}$\,pc$^{-3}$ \citep{Walcher2005, Misgeld_Hilker2011, Norris2014}. They contain stellar masses from 10${^5}$-10${^8}\,M_{\odot}$ in half-light radii of about 1-10\,pc \citep{Georgiev_Boker2014}, and are known to have extended and complex star formation histories \citep[e.g.][]{Walcher2005,Kacharov2018}. 
Nucleation is a common characteristic in galaxies across the Hubble sequence  \citep{Phillips1996, Carollo1998, Boker2002, Boker2004, Cote2006, Turner2012, Georgiev_Boker2014}, with similar occupation fractions for different galaxy cluster environments \citep{Cote2006, Turner2012, denBrok2014, Munoz2015, Sanchez-Janssen2019}. The nucleation fraction peaks close to 100\% in galaxies of $10^9~M_\odot$, and decreases towards both lower and higher galaxy masses \citep[see][for a recent review]{Neumayer2020}.

A large fraction of dwarf galaxies host low-mass NSCs that display characteristics similar to high-mass, metal complex globular clusters (GCs) in the Milky Way. This suggests that such GCs are former NSCs, remnants of the accretion of the host galaxy onto the Milky Way \citep{Zinnecker1988,Boker2008,daCosta2016}. Cosmological merger simulations show that Milky Way-like galaxies should harbour a non-negligible number of these GCs \citep{Pfeffer2014,Kruijssen2018}.  This tidal stripping scenario has been strongly supported by the detection of supermassive black holes in more massive ultra-compact dwarfs (UCDs)  \citep[e.g.][]{Seth2014, Voggel2019}.

Where detailed measurements of kinematics are possible, rotation seems to be a common dynamical ingredient of NSCs \citep{feldmeier2014, Nguyen2018} and is also commonly found in a high fraction of GCs \citep{vandeVen2006, Bellazzini2012, Bianchini2013, Kacharov2014, Fabricius2014, Kimmig2015, Bellini2017, Kamann2018, Bianchini2018, Sollima2019}. The presence of internal rotation can be an indicator of their formation mechanism  \citep{Mastrobuono-Battisti2013, Mastrobuono-Battisti2016, Henault-Brunet2015, Gavagnin2016, Khoperskov2018, Mastrobuono-Battisti2019} and can give important clues on their long-term dynamical evolution \citep[e.g.,][]{Einsel1999, Tiongco2018}. \citet{Fabricius2014} and \cite{Kamann2018} found that GCs with high internal central rotation are more flattened, showing that internal rotation plays an important role in forming the shape of GCs. 

NSC studies focus on galaxies with masses at or above the peak of the nucleation fraction ($10^9~M_\odot$), but no detailed works are available for the low-mass galaxy regime. Thus, low-mass galaxy nucleation is not yet fully understood. In \citet[][hereafter Paper~I]{Alfaro-Cuello2019}, we started a study to understand NSCs in these unexplored low-mass galaxies and the chemo-dynamical properties they display.
We looked for answers to these open questions by studying M54 (NGC~6715), the NSC of the Sagittarius dwarf spheroidal galaxy \citep{Ibata1994} -- hereafter Sgr dSph. This galaxy is currently being disrupted by the tidal field of the Milky Way \citep{Ibata1997}. 
The progenitor of the Sgr dSph galaxy is estimated to have a luminosity of $\sim$10$^8$~L$_\odot$ \citep[$M_v \sim -15$;][]{Niederste-Ostholt2012}, and a dark halo mass with a lower limit of 6$\times10^{10}\,M_\odot$ before infall \citep{Gibbons2017, Mucciarelli2017}. At a distance of 27.6\,kpc \citep{Sollima2009}, M54 is located in the densest region of the Sagittarius stream at the photometric center of its host \citep{Ibata1994, Mucciarelli2017}.

In the literature, the central region of the Sgr dSph galaxy is described as two distinct populations: a metal-poor one with age of $\sim$13~Gyr, identified as M54, and a metal-rich one with age of $\sim$2~Gyr, identified as the nucleus of the Sgr dSph galaxy \citep{Monaco2005a,Siegel2007,Bellazzini2008}.
In Paper~I, we suggested to use the term ``M54" to describe the entire NSC of the Sgr dSph, as in the time of its discovery by Messier in 1778.
We will follow the same approach in this paper, i.e. we will refer to M54 as the entire system, and not just the metal-poor population.

In Paper~I, we presented our large MUSE data set covering an area out to 2.5 times the half-light radius of M54 \citep[R$_{\mathrm{HL}}$$=0\farcm82$,][2010 edition]{Harris1996}, from which we extracted spectra of more than $\sim$6\,600 member stars. From this extensive data set, we characterized the age and metallicity distributions of the stellar populations. We recovered the star formation history of this NSC, detecting (at least) three stellar populations: young metal-rich (YMR), with a mean age of $2.2$\,Gyr and \mbox{[Fe/H]$=-0.04$}; intermediate-age metal-rich (IMR), with mean age of $4.3$\,Gyr and \mbox{[Fe/H]$=-0.29$}; and old metal-poor (OMP), with mean age of $12.2$\,Gyr and \mbox{[Fe/H]$=-1.41$}. 
From our findings we suggested that the OMP population could be a merger remnant of two or more GCs driven to the center of the host galaxy by dynamical friction, thus explaining the large spread in both age and metallicity. 
We note that the OMP population sample might contain old, metal-poor stars that resided in the center of the Sgr dSph galaxy before the infall of the GCs that formed this massive dominant population we observe today. Thus, the large spread in age and metallicity that we measure might be partially due to metal-poor stars from the Sgr dSph. It is not possible to disentangle this with our data.
The first encounter between the Sgr dSph galaxy and the Milky Way disk could have triggered the \textit{in situ} star formation episode of the YMR population in the center of the OMP population, resulting in a more flattened and centrally concentrated structure. The IMR population seems to be part of the Sgr dSph galaxy central field.
However, the three different populations are interacting gravitationally, and thus the IMR cannot be discarded from a dynamical analysis of the system.

\citet{Bellazzini2008} presented a kinematic analysis for a total of $\sim400$ stars, including both metal-poor (OMP) and metal-rich (YMR$~+~$IMR) stars. The stars of both metallicity regimes were found to display consistent  line-of-sight velocities, as suggested by previous studies \citep[e.g][]{Dacosta1995, Ibata1997, Monaco2005b}. 
According to these authors, the metal-rich stars follow a flat velocity dispersion profile of $\sigma=10$~km~s$^{-1}$ within the central $9\arcmin$ radius. The metal-poor population follows a King profile \citep{K66}, which provides a good fit to the surface brightness, at least for the innermost region, as is usually found in GCs. This latter population has a maximum dispersion of $\sigma=14.2$~km~s$^{-1}$ in the center and $\sigma=5.3$~km~s$^{-1}$ at  $\sim3\farcm5$ from the cluster center. Hence, the authors found differences in the kinematics between the two populations in a radius range of $1\farcm5<r<6\farcm5$, suggesting different origins. In this picture, the metal-poor population enabled the collection of the gas that formed the metal-rich population,
resulting in different velocity dispersion profiles. In addition, \citet{Bellazzini2008} found a weak signal of rotation for the metal-poor population $<2$~km~s$^{-1}$. This signal seems to increase at radii larger than $10\arcmin$, where a clear sign of rotation is displayed. Based on line-of-sight velocities, \citet{Baumgardt2018} reported a central velocity dispersion for M54 of $\sigma_0=16.2$~km~s$^{-1}$ \citep[based on data from][]{Bellazzini2008, Ibata2009, Carretta2010b, Watkins2015a}.
Zooming into the center, kinematic observations and dynamical modeling from \citet{Ibata2009} and \citet{Baumgardt2017} suggested the existence of an intermediate-mass black hole (IMBH) of \mbox{M$_{\mathrm{BH}}\sim 10{^4}\,M_{\odot}$}.

Using the exquisite observational data provided by MUSE, we are not just able to use integrated kinematics map of the NSC, as usually done for extragalactic system. For M54, in contrast, we can measure to motion of individual stars, and hence we can obtain the kinematics for each of the different stellar populations in M54. This provides valuable insights into the resulting complex kinematics of NSCs with multiple stellar structures. Since our MUSE data set covers 2.5 times the half-light radius of M54, we are able to evaluate the radial trends of the kinematic quantities up to significant distances from the cluster center.

In this second paper, we present a kinematic analysis of M54 to complement our previous study. We take into account the populations defined in Paper~I, based on age and metallicity information. We present a brief description of the data in Section\,\ref{data}. We describe the method for the kinematic extraction in Section\,\ref{analysis}, including line-of-sight velocity, velocity dispersion and rotation. In Section\,\ref{results} we present the kinematic results for M54 and each of its populations. We present a complement to our analysis using data from \textit{Gaia DR2} to reconstruct the 3D structure in Section~\ref{gaia_analysis}. In Section~\ref{simulations} we show our development on ad hoc $N$-body simulations to support our interpretation of the data and of the formation history of M54. We discuss our findings in Section\,\ref{discussion}, conclude in Section\,\ref{conclu}, and briefly describe the future prospects in Section~\ref{future}.

\section{MUSE Data} \label{data}

As detailed in Paper~I, the data set was acquired with the Multi-Unit Spectroscopic Explorer \citep[MUSE,][]{Bacon2014}, located at the UT4 of the Very Large Telescope at the Paranal Observatory in Chile as part of the programme 095.B-0585(A) (PI: L\"utzgendorf).

The sixteen observed pointings ($59\farcs9 \times 60\farcs0$ each) constitute a 4$\times$4 mosaic covering a contiguous area out to $\sim$2.5 times the half-light radius \citep[$R_\mathrm{HL}=0\farcm82$,][2010 edition]{Harris1996} of M54.
This is a total coverage of $\sim$3\farcm5$\times$3\farcm5, where 1$\arcmin\simeq8$\,pc at an assumed distance of $27.6$\,kpc \citep{Sollima2009}, or \mbox{~28$\times$28\,pc}.
The MUSE data has a wavelength coverage of 4800\,-\,9300\,\AA, with a mean spectral resolution of R$\sim2\,750$ and spatial sampling of $0\farcs2$/pix.

For more details about the data, single stellar extraction and membership estimates see Paper~I.

\section{Analysis}  \label{analysis}

In this section we describe the methods we use for the kinematic extraction that are applied to the samples presented in Section~\ref{results}.

\subsection{Line-of-sight velocity}

\begin{figure}[ht]
\centering
\includegraphics[width=250px]{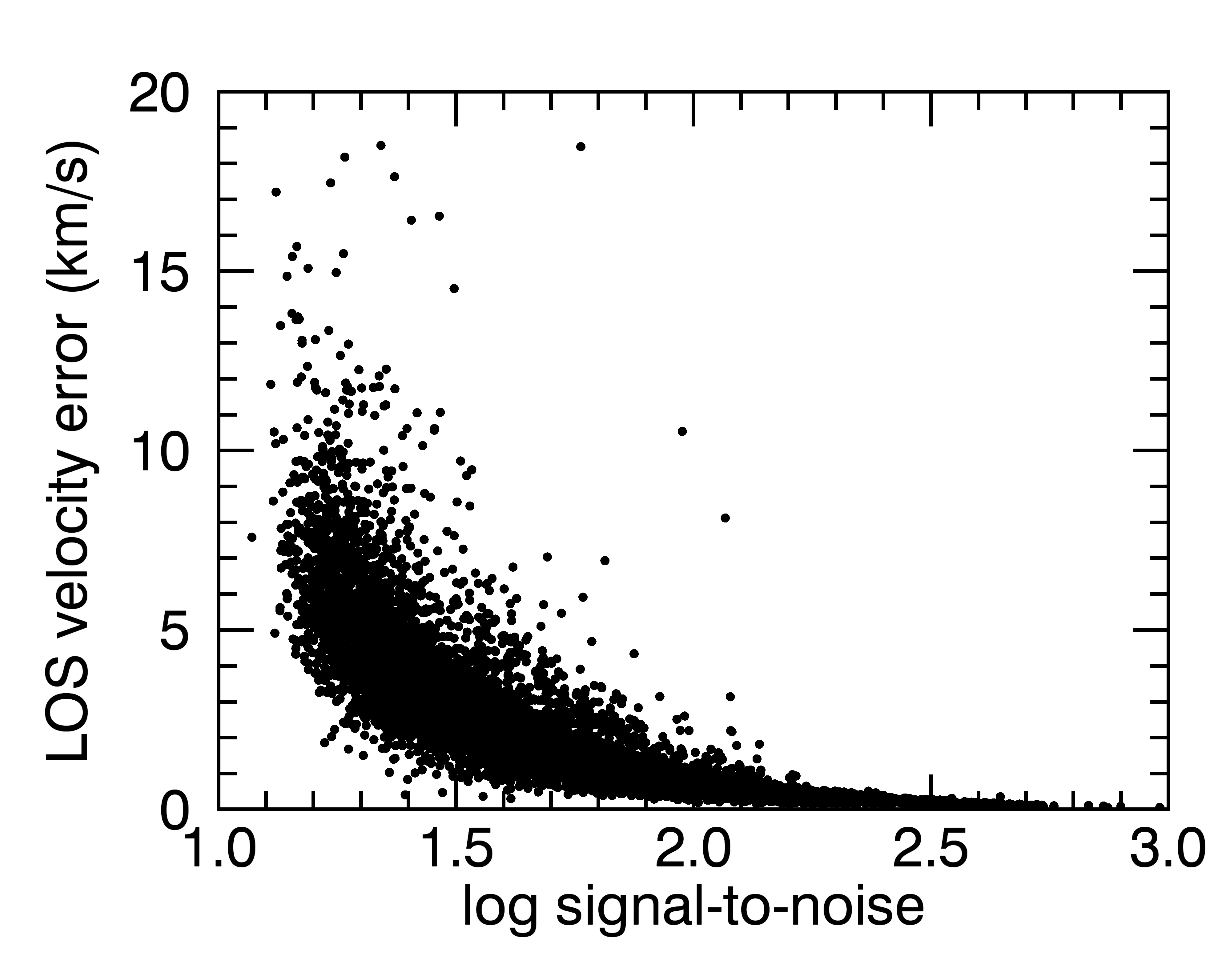}
\caption{Relation between the signal-to-noise logarithm and the velocity errors obtained with UlySS.}
\label{err_vel}
\end{figure}

In Paper~I, we described the determination of physical parameters and line-of-sight (LOS) velocities of the stars using ULySS  \citep[University of Lyon Spectroscopic Analysis Software,][]{Koleva2009}. To obtain the stellar atmospheric parameters, ULySS interpolates and fits to the observed spectra a stellar library of synthetic spectral templates, characterized by metallicity [Fe/H], surface gravity and temperature. We use the synthetic spectroscopic grid, built on the basis of the ELODIE 3.2 library \citep{elodie}.
We present in Figure~\ref{err_vel} the relation between the signal-to-noise logarithm of the stars and the error in the velocity measurements obtained with UlySS. We exclude from the sample all stars with error $>20$~km~s$^{-1}$ leaving 7123 stars. This number includes all extracted stars, we describe the membership cleaning of the sample in Section~\ref{vel_disp_analysis}. The median of the uncertainties of the LOS velocities estimated by UlySS is $2.18$~km~s$^{-1}$.
To test the robustness of our results, we perform the same kinematic analysis for stars with errors lower than $5$~km~s$^{-1}$. This subsample includes 5446 stars and we obtain consistent results.
We corrected the LOS velocity of the stars for the effect of perspective rotation using Equation\,6 in \citet{vandeVen2006}. 

\subsection{Rotation and velocity dispersion} \label{vel_disp_analysis}

Constraining the rotation of the populations of M54 can provide fundamental constraints on their origin.
We measure the rotation and velocity dispersion of M54 simultaneously, using a discrete Bayesian approach similar to the procedure described in \citet{Cordero2017} and \citet{Koch2018} where we consider the LOS velocities of the stars and their respective errors. We approximate our velocity dispersion profile with a Plummer model \citep{Plummer1911}:
\begin{equation}
\sigma(r)^2 = \frac{\sigma^2_0}{\sqrt{1+\frac{r^2}{a^2}}},
\end{equation}
where $\sigma_0$ and $a$ are the central velocity dispersion and the Plummer radius respectively, both treated as free parameters.
The mean LOS velocity ($<V_{\mathrm{LOS}}>$) is also a free parameter.
We approximate the rotation profile with an analytical model \citep{Mackey2013, Kacharov2014, Cordero2017}:
\begin{equation} \label{eq_rotation}
V_{rot}\,\sin{i} = \frac{2V_{max}}{r_{peak}} \times \frac{X_{\mathrm{PA}}}{1+(X_{\mathrm{PA}} / r_{peak})^2},
\end{equation}
where $V_{rot}\,\sin{i}$ is the rotation amplitude at position $X_{\rm PA}$ within a factor of the cluster's inclination ($i$) with respect to the LOS.
The independent variable $X_{\rm PA}$ represents the distance of the different points from the center of M54 along the equatorial axis. 
For this model, we set three free parameters: $V_{max}$, the maximum rotation amplitude at the projected radius $r_{peak}$, and the rotation axis position angle (PA, measured from north $0\degree$ to east $90\degree$). We set a weak Gaussian prior on $r_{peak}$ centered at $0$ with a standard deviation of $4\arcmin$ -- about 5 times the half-light radius and $\sim0.5$ times the tidal radius of the cluster.
We use a Gaussian mixture likelihood model to exclude foreground stars, where the mixing factor is the membership fraction, from which we calculate the membership probability of each star.
To summarize, we account for a total of seven free parameters in our discrete kinematic model: $\sigma_0$, $a$, $<V_{\mathrm{LOS}}>$, $V_{max}$, $r_{peak}$, PA, and membership fraction.
We optimize the Gaussian likelihood function \citep[see Eq. 2 \& 5 in][]{Cordero2017} using the affine-invariant Markov chain Monte Carlo (MCMC) algorithm \textsc{emcee} \citep{Goodman2010, Foreman-Mackey2013}. We use 500 walkers and 5000 iterations to assure all parameters converge. 
We use this method since evaluating discrete data provides more accurate results than other methods that require binning the data, which can end in loss of information.

To be consistent with other studies, we measure the average rotation amplitude ($A_{rot}$) as the half of the difference between the maximum likelihood mean velocity on the two sides of the rotation axis. For the analysis in Section~\ref{rotation_vs_ellipticity}, we additionally measure the rotation at the half-light radius ($A_{\mathrm{HL}}$) and its uncertainty, as the median and standard deviation of the rotation model posterior distribution at that radius.

\section{Kinematic extraction} \label{results}

In this section, we present the results of the kinematic extraction applied to: (i) all M54's member stars as a whole -- with no distinction in populations -- and (ii) the three populations in M54 identified in Paper~I.

The results illustrated in this section are summarized in Table\,\ref{table_kin1}. For an easier comparison, we include in this table some of the population parameters from Paper~I (i.e., [Fe/H], age, ellipticity, etc).
 
\subsection{Kinematics of all M54 member stars }

After excluding background stars from our stellar sample as described in Section~\ref{vel_disp_analysis}, the final sample of M54 member stars includes 6537 stars. These stars occupy a large range in age (0.5 to 14\,Gyr) and metallicity \mbox{($-2.5<\mathrm{[Fe/H]}<0.5$)}. The use of this overall sample mimics the kinematic extraction from a single stellar system, as is commonly the case for more distant NSCs, for which it is impossible to resolve individual stars or to separate different stellar populations.

\subsubsection{Line-of-sight velocity and velocity dispersion} 

We include the LOS velocity map for the resolved stars in the top left panel of Figure\,\ref{voronoi_all}. The dashed black circle shows the half-light radius of M54 \citep[R$_{\mathrm{HL}}$$=0\farcm82$,][2010 edition]{Harris1996}.
For illustration purposes, in the top panel of Figure\,\ref{voronoi_all}, we present Voronoi-binned maps for the mean LOS velocity (middle) and velocity dispersion (right) for our observed M54 member stars. We use the Voronoi tessellation code by \citet{Cappellari2003} as described by \citet{Kamann2018}. We aim for $\sim$45 stars per bin.

Based on our discrete model, the mean V$_{\mathrm{LOS}}$ value is \mbox{$141.54\pm0.21$\,km~s$^{-1}$}. We include in the bottom right panel the best-fitting Plummer profile to the data (solid cyan line), and the $\pm3\sigma$ uncertainty (dashed cyan lines).  The cyan circles show the velocity dispersion from the data points estimated in radial bins of $0\farcm3$, and the vertical dashed line shows the half-light radius of M54. The median of the best-fit central velocity dispersion is \mbox{$\sigma_0=15.91\pm0.33$\,km~s$^{-1}$}.

\subsubsection{Rotation}

\begin{figure*}[ht]
\centering
\includegraphics[width=\textwidth]{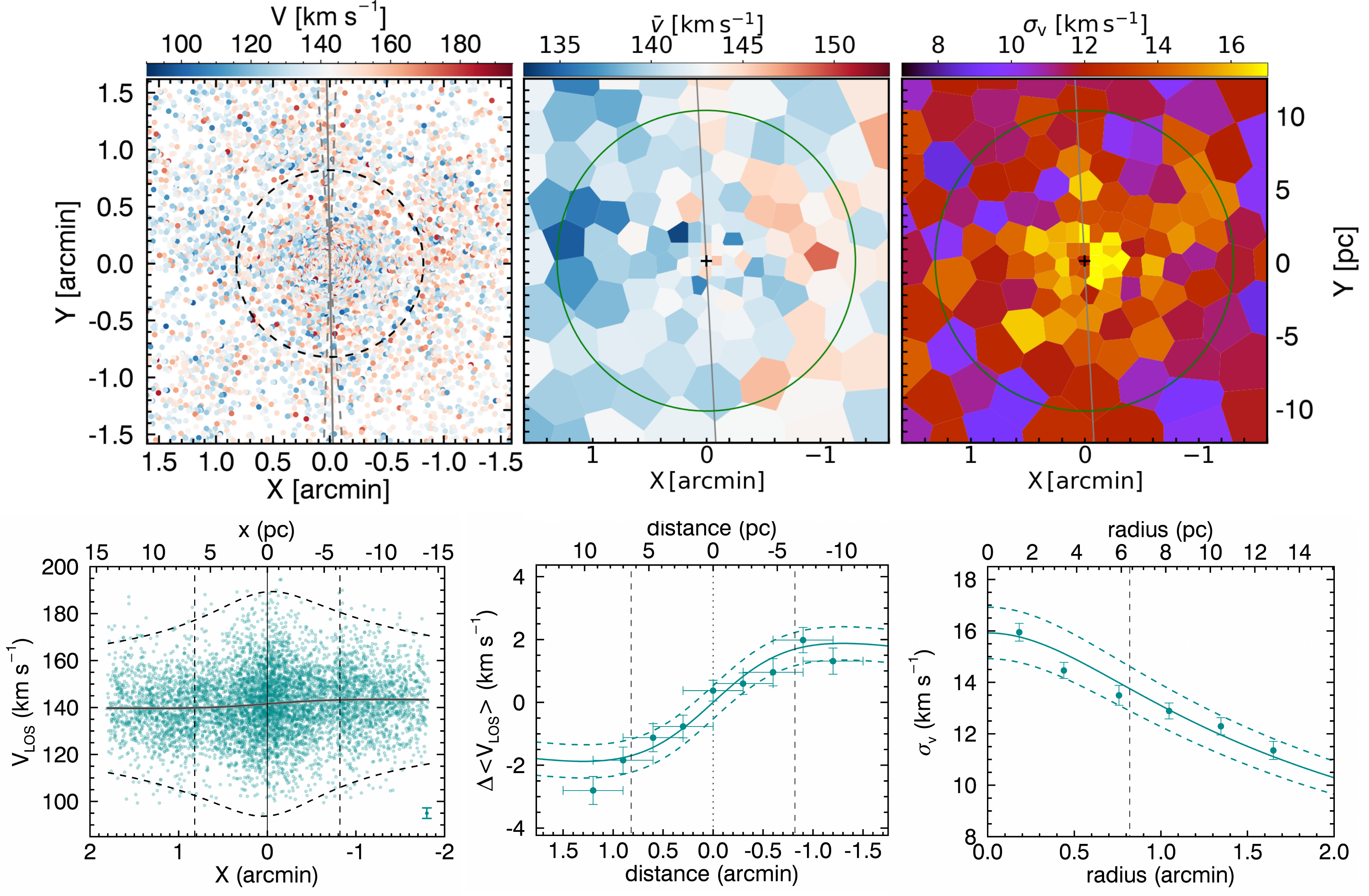}
\caption{Kinematics for all the 6\,537 member stars of M54.  \textbf{Top}: LOS velocity map from resolved stars (left), Voronoi-binned maps for the mean LOS velocity (middle) and velocity dispersion (right).
The overplotted solid gray lines show the median of the best fit rotation axis.
The overplotted dashed black circle in the left panel shows the half-light radius of M54 \citep[R$_{\mathrm{HL}}$$=0\farcm82$,][]{Harris1996}. The green ellipse in the top middle and right panels is overplotted using the half-light radius, position angle and ellipticity estimated in Paper~1 fitting a King profile.
\textbf{Bottom left}: Velocity gradient with respect to a line perpendicular to the rotation axis. The solid black line shows the median of the best-fit rotation models. The dashed black lines show $\pm3$ times the velocity dispersion. \textbf{Bottom middle}: Rotation profile for M54 member stars (solid cyan line) obtained as the median of the best-fit rotation models using Eq.\,2. The cyan circles represent the rotation profile derived as the difference between the median velocity and the systemic velocity for overlapping bins of $0\farcm6$ along the line perpendicular to the rotation axis. The horizontal and vertical error bars represent the radial bin size and the uncertainties in the offset of the median velocity, respectively.  \textbf{Bottom right}: Velocity dispersion profile for the M54 members obtained as the median of the best-fit Plummer profiles (solid cyan line). Dashed cyan lines show the $\pm3\sigma$ uncertainty for each best-fit model. 
The cyan circles show the velocity dispersion from the data points estimated in radial bins of $0\farcm3$.
The vertical dashed lines in both panels show the half-light radius of M54 \citep[R$_{\mathrm{HL}}$$=0\farcm82$,][2010 edition]{Harris1996}. }
\label{voronoi_all}
\end{figure*}

The results of the rotation analysis for M54 are presented in the bottom left and middle panels of Figure\,\ref{voronoi_all}. The left panel shows the velocity gradient with respect to a line perpendicular to the rotation axis. The solid black line is the median of the best-fit rotation models. The dashed black lines show $\pm3$ times the velocity dispersion.
In the bottom middle panel of Figure\,\ref{voronoi_all} we show the best-fit rotation models to the discrete data obtained using Eq.~\ref{eq_rotation}, as a solid cyan line, and the $\pm3\sigma$ uncertainty as dashed cyan lines. 
For representation, the cyan circles represent the rotation profile derived as the difference between the median velocity and the systemic velocity for overlapping bins of $0\farcm6$ along the line perpendicular to the rotation axis. The horizontal and vertical error bars represent the radial bin size and the uncertainties in the offset of the median velocity, respectively.
The vertical dashed lines show the half-light radius of M54 \citep[R$_{\mathrm{HL}}$$=0\farcm82$,][2010 edition]{Harris1996}. The best-fit rotation axis corresponds to $2.9\degree\pm8.8\degree$. We detect a low -- but still considerable -- amount of rotation in M54. We obtain a rotation amplitude of \mbox{$A_{rot}=1.16\pm0.36$\,km~s$^{-1}$}, and a maximum rotation of \mbox{$V_{max}=1.88\pm0.70$\,km~s$^{-1}$}.  

\subsection{Kinematics of M54's populations}

In this section, we study the kinematics of each of the three populations identified via age and metallicity in Paper~I: YMR (630 stars), IMR (649 stars), and OMP (2550 stars).
For better kinematic characterization, we do not exclude stars with ages with relative errors $>40$\%, thus, the number of stars for each population is slightly larger than those considered in Paper~I.
The number of stars in the full M54 sample is larger than the sum of stars in the three populations, because it also includes horizontal branch stars ($\sim$650) and stars that could not clearly be attributed to either of the three subpopulations when we created the Multi-Gaussian model in Paper~I.

\subsubsection{Line-of-sight velocity and velocity dispersion}

\begin{figure*}[ht]
\centering
\includegraphics[width=\textwidth]{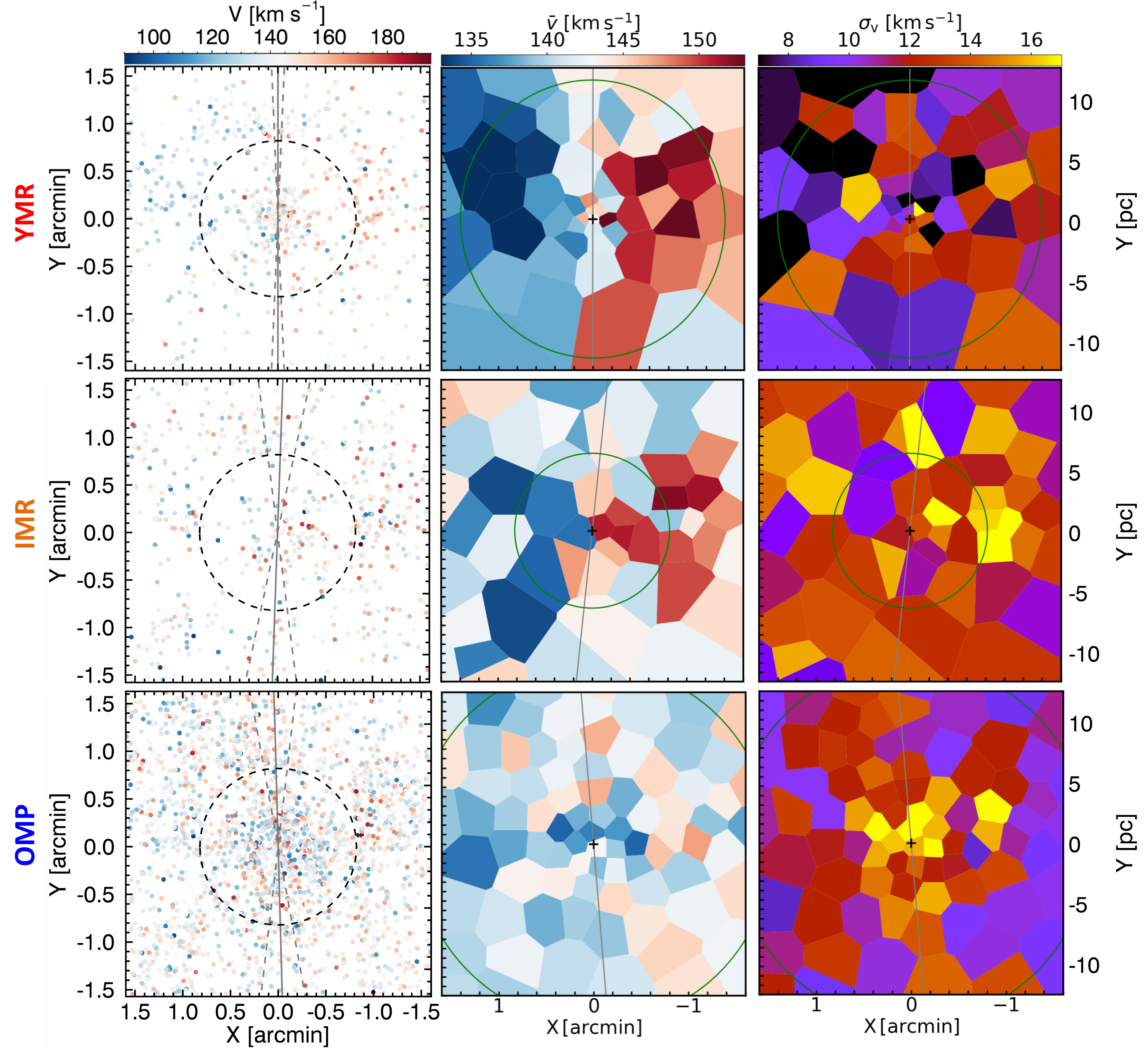}
\caption{Kinematic extraction for the three M54 populations: YMR in top panels, IMR in middle panels, and OMP in bottom panels. The left panel shows the LOS velocity map for the stars in each population. The middle and right panels show the Voronoi-binned maps for the mean LOS velocity and velocity dispersion, respectively. Overplotted dashed black circles in the left panels show the half-light radius of M54 \citep[R$_{\mathrm{HL}}$$=0\farcm82$,][]{Harris1996}. 
The green ellipses in the middle and right panels are overplotted using the half-light radius, position angle and ellipticity estimated in Paper~1 for a King profile. Since these values were not obtained for the IMR population, we overplotted a spherical shape.
The overplotted gray solid lines show the median of the best-fit rotation axis.}
\label{voronoi_3pop}
\end{figure*}

In the left, middle and right panels of Figure\,\ref{voronoi_3pop}, we present the LOS velocity maps of the resolved stars, and for representation, the Voronoi-binned velocity and velocity dispersion maps for each of the populations in M54: YMR, IMR, and OMP, from top to bottom, respectively. For the Voronoi maps, we set a target number of stars per bin depending of the size of the sample of the populations: 25 for the OMP, and 7 for the IMR and YMR populations. 

We found very similar mean LOS velocities for the three populations. The mean velocities are \mbox{$142.01\pm0.52$}\,km~s$^{-1}$ for the YMR, \mbox{$142.61\pm0.59$}\,km~s$^{-1}$ for the IMR, and \mbox{$141.22\pm0.27$}\,km~s$^{-1}$ for the OMP. 
The LOS velocities are consistent when only considering stars with errors lower than $5$~km~s$^{-1}$, showing variations of $\sim0.3$~km~s$^{-1}$.
From the Voronoi-binned LOS velocity maps we observe a clear velocity gradient for the YMR population.

We present in the right panels of Figure~\ref{best_fits} the velocity dispersion profile for each population: YMR (top), IMR (middle), and OMP (bottom). The profile is obtained as the median of the best-fit Plummer models to the discrete data. The dashed lines show the $\pm3\sigma$ uncertainty for each profile.
For representation, in each case, the colored circles show the velocity dispersion from the data points estimated in radial bins of $0\farcm3$.
The vertical dashed lines in both panels show the half-light radius of M54 \citep[R$_{\mathrm{HL}}$$=0\farcm82$,][2010 edition]{Harris1996}.
We estimate a central velocity dispersion of \mbox{$\sigma_0=12.56\pm0.72$\,km~s$^{-1}$} for the YMR, and \mbox{$\sigma_0=15.21\pm0.89$\,km~s$^{-1}$} for the IMR. For the YMR and IMR velocity dispersion profiles we observe that they rather follow a flat distribution along the observed radius, thus the best-fit Plummer profiles do not provide a good description of the data for these two populations. Consistently, the best fit velocity dispersion profiles for the YMR and IMR populations are flat when we only consider stars with errors lower than $5$~km~s$^{-1}$, displaying slightly flatter profiles with central $\sigma$ values lower by $\sim1$~km~s$^{-1}$.
In contrast, the OMP velocity dispersion profile (blue) is very well fit by a Plummer model. The central velocity dispersion for the OMP population is \mbox{$\sigma_0=15.30\pm0.54$\,km~s$^{-1}$}.

\subsubsection{Rotation}

\begin{figure*}
\centering
\includegraphics[width=\textwidth]{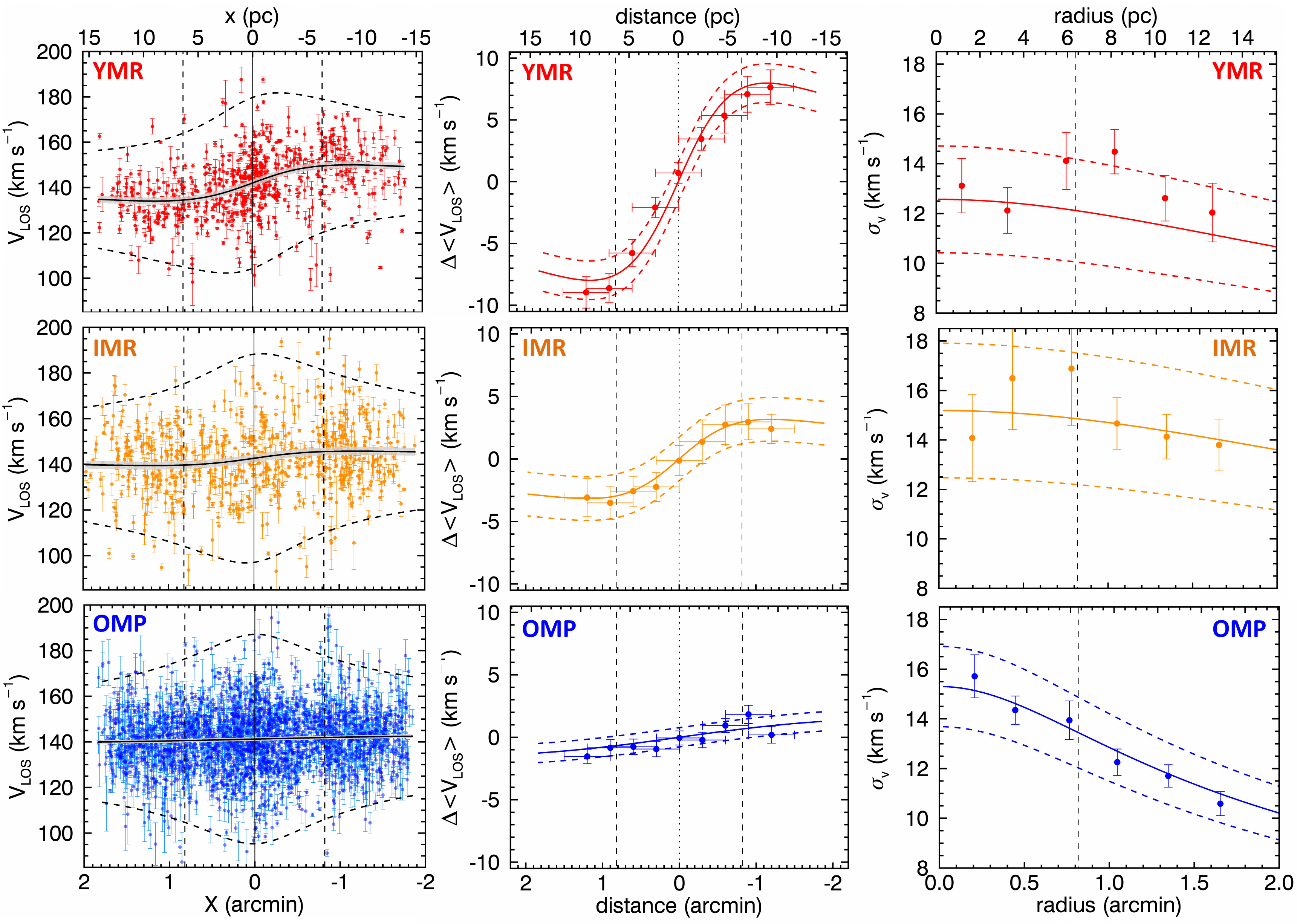}
\caption{\textbf{Left panels}: Velocity gradient with respect to a line perpendicular to the rotation axis for the three populations: YMR (top), IMR (middle), and OMP (top). The solid black line shows the median of the best-fit rotation models and the shaded area corresponds to $\pm3\sigma$. The dashed black lines show $\pm3$ times the velocity dispersion. 
\textbf{Middle panels:} Rotation profile for the three populations obtained as the median of the best-fit rotation models using Eq.\,2. The dashed lines show the $\pm3\sigma$. The colored circles represent the rotation profile derived as the difference between the median velocity and the systemic velocity for overlapping bins of $0\farcm6$ along the line perpendicular to the rotation axis. The horizontal and vertical error bars represent the radial bin size and the uncertainties in the offset of the median velocity, respectively. 
\textbf{Right panels}: Velocity dispersion profile for the three populations obtained as the median of the best-fit Plummer profiles (solid lines).
In all panels, the vertical dashed lines show the half-light radius of M54 \citep[R$_{\mathrm{HL}}$$=0\farcm82$,][2010 edition]{Harris1996}.}
\label{best_fits}
\end{figure*}

In the left panels of Figure~\ref{best_fits}, we present the velocity gradient with respect to a line perpendicular to the rotation axis for the three populations: YMR (top), IMR (middle), and OMP (bottom). The solid black line shows the median of the best-fit rotation models. The dashed black lines show $\pm3$ times the velocity dispersion.
In the middle panels, we present the median of the best-fit rotation profiles obtained using Eq.\,2 as solid lines for the three populations. The dashed lines of the same colors represent the $\pm3\sigma$ uncertainties.
The vertical dashed lines show the half-light radius of M54 \citep[R$_{\mathrm{HL}}$$=0\farcm82$,][2010 edition]{Harris1996}.
As an illustration, the colored circles represent the rotation profile derived as the difference between the median velocity and the systemic velocity for overlapping bins of $0\farcm6$ along the line perpendicular to the rotation axis. The horizontal and vertical error bars represent the radial bin size and the uncertainties in the offset of the median velocity, respectively. 

The highest maximum rotation is displayed by the YMR population with $V_{max}=7.98\pm1.01$\,km~s$^{-1}$, with a median rotation axis at \mbox{$-0.1\degree\pm7.2\degree$}. The second highest maximum rotation is displayed by the IMR population with a median of $V_{max}=3.15\pm1.95$\,km~s$^{-1}$, less than half that of the YMR population, and a median rotation axis at \mbox{$-5.9\degree\pm26.4\degree$}.
The OMP population shows a low amount of rotation, a best-fit value of maximum rotation of $V_{max}=1.58\pm1.14$\,km~s$^{-1}$, with a rotation axis at \mbox{$4.3\degree\pm23.7\degree$}.
In order to compare with other studies, we also compute the average amplitudes of rotation\footnote{As half the difference of the mean velocities on the two sides of the rotation axis} of the three populations as follows: \mbox{$A_{rot}(\rm OMP)=0.75\pm0.55$\,km~s$^{-1}$}, \mbox{$A_{rot}(\rm IMR)=2.26\pm0.16$\,km~s$^{-1}$}, and the fastest rotator with \mbox{$A_{rot}(\rm YMR)=4.91\pm0.90$\,km~s$^{-1}$}.
The differences in the velocity dispersion and rotation properties suggest different origins for these populations.

\begin{deluxetable*}{c|cccc}
\tablecaption{Summary of the observed properties of M54 and its stellar populations.  \label{table_kin1}}
\tablehead{
\colhead{populations} & \colhead{M54 members} &   \colhead{YMR} & \colhead{IMR} & \colhead{OMP} }
\startdata
  [Fe/H]                          & $-2.5$\,-\,0.5   & $-0.04\pm0.01^a$   & $-0.29\pm0.01^a$    & $-1.41\pm0.01^a$ \\
  $\sigma_{\mathrm{[Fe/H]}}$      & -                & $0.12\pm0.01^a$    & $0.16\pm0.01^a$     & $0.24\pm0.01^a$ \\
  Age (Gyr) $^a$                  & 0.5\,-\,14       & $2.16\pm0.03^a$    & $4.28\pm0.09^a$     & $12.16\pm0.05^a$ \\ 
  $\sigma_{\mathrm{Age}}$ (Gyr)   & -                & $0.20\pm0.03^a$    & $1.16\pm0.07^a$     & $0.92\pm0.04^a$  \\ 
  $\epsilon$                      & $0.13\pm0.03$    & $0.31\pm0.10^a$    &  $^b$               & $0.16\pm0.06^a$     \\
  $\epsilon_{\mathrm{HL}}$        & $0.08\pm0.04$    & $0.15\pm0.13$      &  $^b$               & $0.09\pm0.06$     \\ 
  Position Angle (deg)            & $8.83\pm6.62$    & $4.56\pm10.84^a$   &  $^b$               & $16.83\pm12.81^a$  \\
  Median V$_{\mathrm{LOS}}$ (km~s$^{-1}$) & $141.54\pm0.21$  & $142.01\pm0.52$  & $142.61\pm0.59$   & $141.22\pm0.27$  \\
  $\sigma_0$ (km~s$^{-1}$)        & $15.91\pm0.33$ & $12.56\pm0.72$     & $15.21\pm0.89$      & $15.30\pm0.54$     \\
  $A_{rot}$ (km~s$^{-1}$)         & $1.16\pm0.36$    & $4.91\pm0.90$      & $2.26\pm1.24$       & $0.75\pm0.55$      \\
  $A_{\mathrm{HL}}$ (km~s$^{-1}$) & $1.60\pm0.29$    & $7.30\pm0.84$      & $2.00\pm0.89$       & $0.73\pm0.38$      \\
  $V_{max}$ (km~s$^{-1}$)         & $1.88\pm0.70$    & $7.98\pm1.01$      & $3.15\pm1.95$       & $1.58\pm1.14$      \\
  Rotation axis (deg)             & $2.9\pm8.8$      & $-0.1\pm7.2$       & $-5.9\pm26.4$       & $4.3\pm23.7$   \\
  Number of stars                 & $6537$           & $630$             & $649$              & $2550$            \\ \hline
\enddata
\tablecomments{~ $^a$: Values from Paper~I.  $^b$: Measurements which did not converge to a value. $\sigma_{\mathrm{[Fe/H]}}$ and $\sigma_{\mathrm{Age}}$ correspond to the [Fe/H] and age intrinsic spreads, respectively. $\epsilon$: ellipticity estimated at the entire FOV. $\epsilon_{\mathrm{HL}}$: ellipticity estimated at the half-light radius. $V_{\mathrm{LOS}}$: line-of-sight velocity. $\sigma_0$: central velocity dispersion. $A_{rot}$: average rotation amplitude. $A_{\mathrm{HL}}$: rotation at the half-light radius. $V_{max}$: maximum rotation. \textit{Note}: The number of stars in the M54 sample is larger than the sum of the three populations because it includes: horizontal branch stars (which were excluded for the age estimates), and tars that could not clearly be attributed to either of the three subpopulations when we created the Multi-Gaussian model in Paper~I.}
\end{deluxetable*}

\subsection{V/$\sigma_0$ versus $\epsilon$} \label{rotation_vs_ellipticity}

\begin{figure}
\centering
\includegraphics[width=245px]{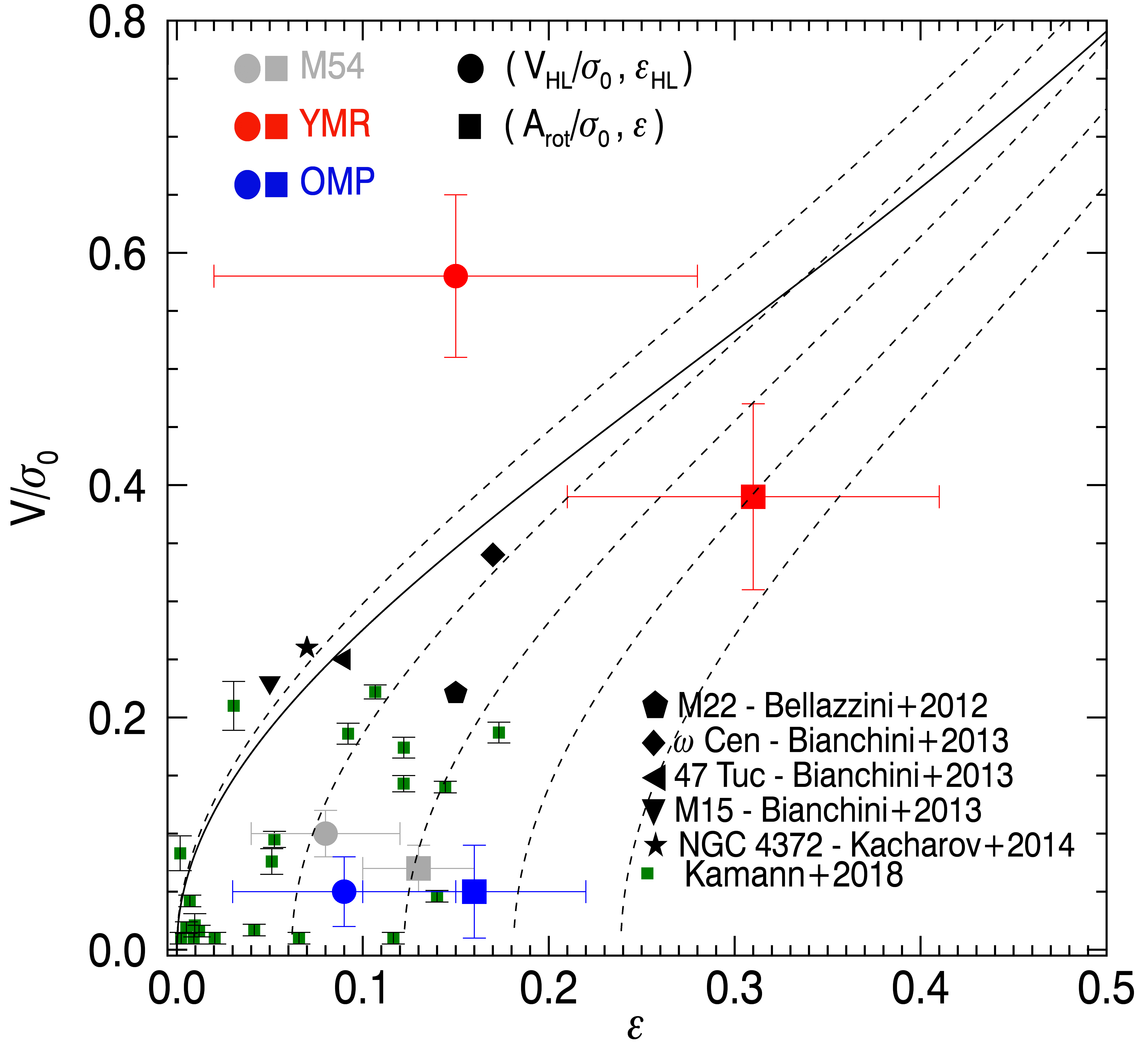}
\caption{($V/\sigma_0,\epsilon$) diagram.
We present our estimates for the entire sample of stars (M54), the YMR and OMP populations in gray, red and blue, respectively. The circles show the estimates based on the rotation and ellipticipy at the half-light radius ($V_{\mathrm{HL}}/\sigma_0,\epsilon_{\mathrm{HL}}$), the squares, on the rotation amplitude and ellipticity over the entire FoV ($A_{rot}/\sigma_0,\epsilon$).
For comparison, we include values for GCs from the literature \citep[e.g.][]{Bellazzini2012, Bianchini2013, Kacharov2014} and in green squares the GC sample from \citet{Kamann2018}. The solid black line gives the relation for an isotropic oblate rotator as \citet{Binney2005}. The dashed black lines show the relation for oblate rotators with anisotropy parameters of \mbox{$\delta=0.05,0.1,0.15,0.2$}, from left to right, as \citet{Cappellari2007}.}
\label{diagnostic}
\end{figure}

The ratio between the rotation and central velocity dispersion versus ellipticity  (V/$\sigma_0$,$\epsilon$) diagrams have been introduced and are widely used to evaluate how the rotation affects the shape of galaxies \citep[e.g.,][]{Binney2005, Cappellari2007, Emsellem2011}. However, in the last few years these diagrams have also been applied to GCs \citep[e.g][]{Bellazzini2012, Bianchini2013, Kacharov2014, Fabricius2014, Lardo2015, Kimmig2015, Kamann2018, Bianchini2018}. 
In this approach, several difficulties arise for both clusters and galaxies, because: (i) the rotation measured in clusters depends on the inclination angle of the rotation axis, which is measurable only when 3D kinematics are available (LOS velocities and proper motions, e.g., \citealt{Bianchini2018,Sollima2019}); (ii) both ellipticity and rotation change as a function of radius \citep[e.g.,][]{Geyer1983}; and (iii) the anisotropy also varies with radius \citep[e.g.,][]{vandeVen2006,Jindal2019}. In spite of these issues, (V/$\sigma_0$,$\epsilon$) diagrams still provide a good first diagnostic of the role of rotation on the shape of these stellar systems, e.g., how flattened they are \citep{Fabricius2014, Kamann2018}. 
In Figure\,\ref{diagnostic} we present a (V/$\sigma_0$,$\epsilon$) diagram where we add the measurements for the whole M54 sample (gray), the YMR (red), and OMP (blue) populations. We do not include the IMR population in this section since the method used to obtain the 2D morphology parameters did not converge to an ellipticity value. 
Since different definitions of $V/\sigma_0$ are used in the literature, we decide to compute the $V/\sigma_0$ in two different ways: (i) based on the rotation and ellipticity at the half-light radius ($V_{\mathrm{HL}}/\sigma_0,\epsilon_{\mathrm{HL}}$) and represented by circles, and (ii) based on the rotation amplitude and ellipticity over the entire FOV ($A_{rot}/\sigma_0,\epsilon$), represented by squares. 
The ellipticity at the half-light radius is derived using the method described in Paper~I used to estimate the ellipticity across the entire FoV.
We explore the possibility that the inclination angle, unknown a priori, could be different from an edge-on view and thus introduce a correction for inclination as described in \citet{Cappellari2007}. We present the respective corrected values in Table~\ref{table_diagnostic}. We will show in Section~\ref{gaia_analysis} that by combining LOS data and \textit{Gaia DR2} proper motions we can estimate that the most likely inclination angle is $\sim90\degree$ (egde-on view). However, inclination angles within $60\degree$ and $90\degree$ cannot be excluded within the $1\sigma$ uncertainties. Therefore we illustrate the effects of a $60\degree$ inclination angle, and consider the corrected value as a conservative (upper limit) estimate for the intrinsic  ($V/\sigma_0,\epsilon$). 

\begin{deluxetable}{c|cccc}
\tablecaption{Corrected $(V\sigma)$ and ellipticity ($\epsilon$) values \label{table_diagnostic}}
\tablehead{
\colhead{} & \colhead{M54} &   \colhead{YMR} & \colhead{OMP}}
\startdata
  $A_{rot}/\sigma_0$          & 0.08 & 0.44 & 0.06    \\
  $\epsilon$                  & 0.18 & 0.45 & 0.22  \\   
  $V_{\mathrm{HL}}/\sigma_0$  & 0.11 & 0.70 & 0.05  \\   
  $\epsilon_{HL}$             & 0.11 & 0.21 & 0.12 \\   \hline
\enddata
\tablecomments{All values were obtained correcting for inclination as described in \citet{Cappellari2007}. 
}
\end{deluxetable}

In Figure\,\ref{diagnostic}, we include for comparison measurements for GCs from \citet{Bellazzini2012, Bianchini2013, Kacharov2014, Kamann2018}, who consistently measure ($V/\sigma_0$) at the half-light radius or beyond.
We overplot as a solid black line the relation for an isotropic oblate rotator as in \citet{Binney2005}, and the dashed black lines show the relation for oblate rotators with the global anisotropy $\delta$ \citep[see][]{Binney_Tremaine1987} parameters \mbox{$\delta=0.05,0.1,0.15,0.2$} from left to right, as presented in \citet{Cappellari2007}.

For the OMP population we observe a low amount of rotation, with only minor differences when estimating this value at the half-light radius R$_{\mathrm{HL}}$ (blue circle) or over the entire FOV of the sample (blue square). 
In spite of a low amount of rotation, the ellipticity estimated over the entire FOV shows that this population is flattened, similar to fast rotating GCs such as the cases plotted in Figure~\ref{diagnostic}: $\omega$~Cen (black diamond) and M22 (black pentagon). We will discuss a possible explanation for this in Section~\ref{omp+ymr}. 

Our rotation estimates for M54 and its OMP population are consistent with the estimate for the metal-poor population by \citet{Bellazzini2012}, who considered an ellipticity of $\epsilon=0.06$ from \citet[][2010 edition]{Harris1996}. A more consistent agreement between the rotation and ellipticity of the OMP population reported by \citet{Bellazzini2012} was expected since their sample does not include young and metal-rich stars.

\newpage

\subsection{Energy equipartition} \label{energy_eq}

In this section we explore the possibility that differences in the kinematics could arise from dynamical relaxation processes, namely energy equipartition, and not necessarily only from different formation mechanisms of the different populations.

Dynamical interactions between stars of different masses naturally bring a stellar system into a mass-segregated configuration, with massive stars losing energy and sinking to the centre. In parallel, stars with different masses will display different kinematics, a process known as partial energy equipartition \citep[see e.g.,][]{Spitzer1969, Trenti_vanderMarel2013, Bianchini2016} that acts on the timescale of a few relaxation times. In particular, massive stars are expected to have a lower velocity dispersion than low-mass stars, with a velocity dispersion ratio that depends on the relaxation state of the cluster and the stellar mass difference between the populations.
In this study, the YMR population is characterized by an average stellar mass\footnote{We obtained the stellar mass for each star from the best-fit isochrone as described in Paper~1.} of $1.53~M_\odot$, while that of the OMP population is a factor $\sim2$ smaller, $0.82~M_\odot$. Therefore, different  kinematics could arise during their $\sim2$~Gyr coevolution (i.e., the age of the YMR population).

Following Eq.~3 of \citet{Bianchini2016}, the ratio between the velocity dispersions of two stellar components due to partial energy equipartition is given by:

\begin{equation} \label{en_eq}
    \frac{\sigma_\mathrm{OMP}}{\sigma_\mathrm{YMR}} = \mathrm{exp}\left(-\frac{1}{2} \frac{m_{OMP}-m_{YMR}}{m_{eq}}\right)
\end{equation}

with $m_{YMR}$ and $m_{OMP}$ the typical masses of the YMR and OMP component and $m_{eq}$ the equipartition parameter quantifying what degree of energy equipartition the system was able to reach (i.e., how strong the kinematic differences are expected to be). While the equipartition parameter is not known a priori, it strongly depends on the relaxation state of a stellar system: clusters that have lived many relaxation times are characterized by a higher degree of energy equipartition. 

We can use Eq.~6 of \citet{Bianchini2016} to estimate the value of $m_{eq}$, starting from the current core relaxation time of M54 given by \citet{Harris1996}, T$_{rc}=1.738\times10^8$~yr. This implies that M54 has dynamically evolved for $\sim12$ relaxation times, which yields $m_{eq}=2.03~M_\odot$. Using this value in Eq.~\ref{en_eq}, we estimate that dynamical processes can produce a ratio between the velocity dispersion of the YMR and OMP population of $\sigma_\mathrm{OMP}/\sigma_\mathrm{YMR}=1.19$. This difference is fully consistent with the one we observe ($\sim1.22$). 

In summary, the effect of partial energy equipartition due to dynamical relaxation can indeed explain the kinematic differences between the YMR and OMP population. In other words, the two populations need not necessarily have formed with different velocity dispersions in order to display the currently observed differences. We caution, however, that further studies of the interplay between energy equipartition and rotation are needed for a comprehensive interpretation of the dynamical evolution of the kinematic differences.

\section{M54 kinematics with \textit{Gaia DR2}} \label{gaia_analysis}

The diagnostic diagram $(V/\sigma_0,\epsilon)$ suffers from the degeneracy of information of the inclination angle, i.e., unknown intrinsic amount of angular momentum. This limitation can be compensated using 3D velocity data from \textit{Gaia DR2}.
The recent availability of precision astrometry from \textit{Gaia} Data Release 2 \citep[DR2,][]{GaiaCollaboration2016, GaiaCollaboration2018} opens the possibility of studying the kinematics of GCs in 3D  \citep[e.g.,][]{Sollima2019}. Here we exploit for the first time the available \textit{Gaia} proper motions with the goal of characterizing the intrinsic dynamical properties of M54's populations.

We cross-match our MUSE M54 stellar sample with the \textit{Gaia DR2} catalog \citep{GaiaCollaboration2016, GaiaCollaboration2018} finding a total of 638 OMP stars, 109 YMR stars, and 32 IMR stars, for which the full 3D velocity vector is measured. These stars have magnitudes $\mathrm{G}<20$, and are within $\sim2\farcm0$ from M54's center.
However, M54 represents a challenging environment for \textit{Gaia} astrometry due to its relatively large distance and the resulting high degree of crowding. To obtain reliable measurements, we need to further restrict the sample to only stars with high-quality proper motion measurements. For this purpose, we perform quality cuts, following some of the procedures illustrated in \citet{Lindegren2018} and \citet{Vasiliev2019}, to eliminate stars with bad astrometric measurements and those strongly affected by crowding. These cuts, based on parameters provided in the \textit{Gaia DR2} catalog\footnote{For a detailed description of the \textit{Gaia} parameters see \citealp{Lindegren2018}.}, include:
\begin{itemize}
\item $\texttt{astrometric\_gof\_al}<0.5$
\item $\texttt{astrometric\_excess\_noise}<1$
\item unit weight error, $uw=<1.2$ (see \citealp{Lindegren2018} for definition)
\item $\texttt{phot\_bp\_rp\_excess\_factor}<2.0 + 0.06\,(\texttt{bp-rp})^2$, with \texttt{bp-rp} the color in \textit{Gaia} filters.
\item proper motions errors $<0.5$ mas yr$^{-1}$ (corresponding to $\sim60$ km s$^{-1}$ at M54's distance).
\end{itemize}

Our final sample consists of a total of 108 OMP stars and 15 from the YMR population, with an average uncertainty in proper motion measurements of 0.19\,mas~yr$^{-1}$ (corresponding to $\sim$25\,km~s$^{-1}$). The final sample for the IMR populations consist of two stars only. Since this is not sufficient to perform the analysis, we will just consider the YMR and OMP populations.

\subsection{Rotation with \textit{Gaia DR2}}

We transform the positions and velocities from celestial to Cartesian coordinates using Eq.\,2 of \citet{GaiaCollaboration2018} and Eq.\,1 of \citet{vandeVen2006} (see also \citealt{Bianchini2018}). 
We correct the LOS velocities and proper motions for perspective rotation following Eq.\,6 in \citet{vandeVen2006}. Finally, when converting proper motions from mas~yr$^{-1}$ to km~s$^{-1}$, we assume a distance of 27.6~kpc \citep{Sollima2009}. As a further test, we repeat the analysis using a distance of 26.5~kpc \citep[][2010 edition]{Harris1996} and 28.4~kpc \citep{Siegel2011}, obtaining consistent results.

We measure the mean motions of the two proper motion components using the likelihood employed in \citet[][Eq. 2 and 3]{Bianchini2018}. For the OMP stars we obtain mean motions of \mbox{$(\mu_{x},\mu_{y})=(2.80\pm0.03,1.40\pm0.02)$~mas~yr$^{-1}$}, and \mbox{$(\mu_{x},\mu_{y})=(2.77\pm0.09,1.48\pm0.08)$~mas~yr$^{-1}$} for the YMR stars. These values are consistent with each other and with the value reported by \citet{Vasiliev2019}, who made no distinction between populations in M54. Together with the fact that the mean LOS velocities of the OMP and YMR components are also consistent with each other, this further indicates that the two stellar populations are comoving and therefore they belong to the same bound stellar system.

To search for a signature of rotation we consider polar coordinates in the plane of the sky and analyze the tangential component of proper motions ($\mu_t$). 
Both populations show a mean value of $\mu_t$ consistent with zero ($\mu_{t,OMP}=0.012\pm0.030$~mas~yr$^{-1}$ and $\mu_{t,YMR}=-0.058\pm0.113$~mas~yr$^{-1}$), indicating no signal of rotation, within their 1$\sigma$ uncertainties. However, we note that the sample size of the YMR population is composed of only 15 stars and, moreover, the putative rotation signal that we are trying to measure ($<10$\,km~s$^{-1}$) is below the nominal systematic uncertainties for \textit{Gaia DR2} data ($0.07$~mas~yr$^{-1}$, \citealp{Lindegren2018}). Therefore a presence of rotation cannot be excluded with this analysis and the current data.

\subsection{Rotation from 3D kinematics} 

Since our sample consists of the full three-dimensional velocity vectors, we can estimate the intrinsic rotation exploiting simultaneously the three velocity components, following the likelihood method described by \citet[][likelihood in Eq.\,3]{Sollima2019}. We assume a constant rotation amplitude\footnote{This is a first order approximation suitable for the low number of data points available.} within the cluster and take into consideration the discrete velocity measurements, their uncertainties and their covariance matrix. We sample the likelihood and derive the $1\sigma$ error using \texttt{emcee} by \citet{Foreman-Mackey2013} keeping as free parameters the position angle of the rotation axis in the plane of the sky ($\theta_0$, measured from west of north), the inclination angle of the rotation axis with respect to the LOS ($i$, with $i=\pi/2$, corresponding to an edge-on view), and the amplitude of the intrinsic rotation ($A_{rot}$, measured in km~s$^{-1}$). We fix a velocity dispersion of 13\,km~s$^{-1}$ (as derived globally from LOS velocities only) for the modeling. We repeated the analysis using different values between $10$ and $15$~km~s$^{-1}$ and saw no differences in the final results.

\begin{figure*}
\centering
\includegraphics[width=480px]{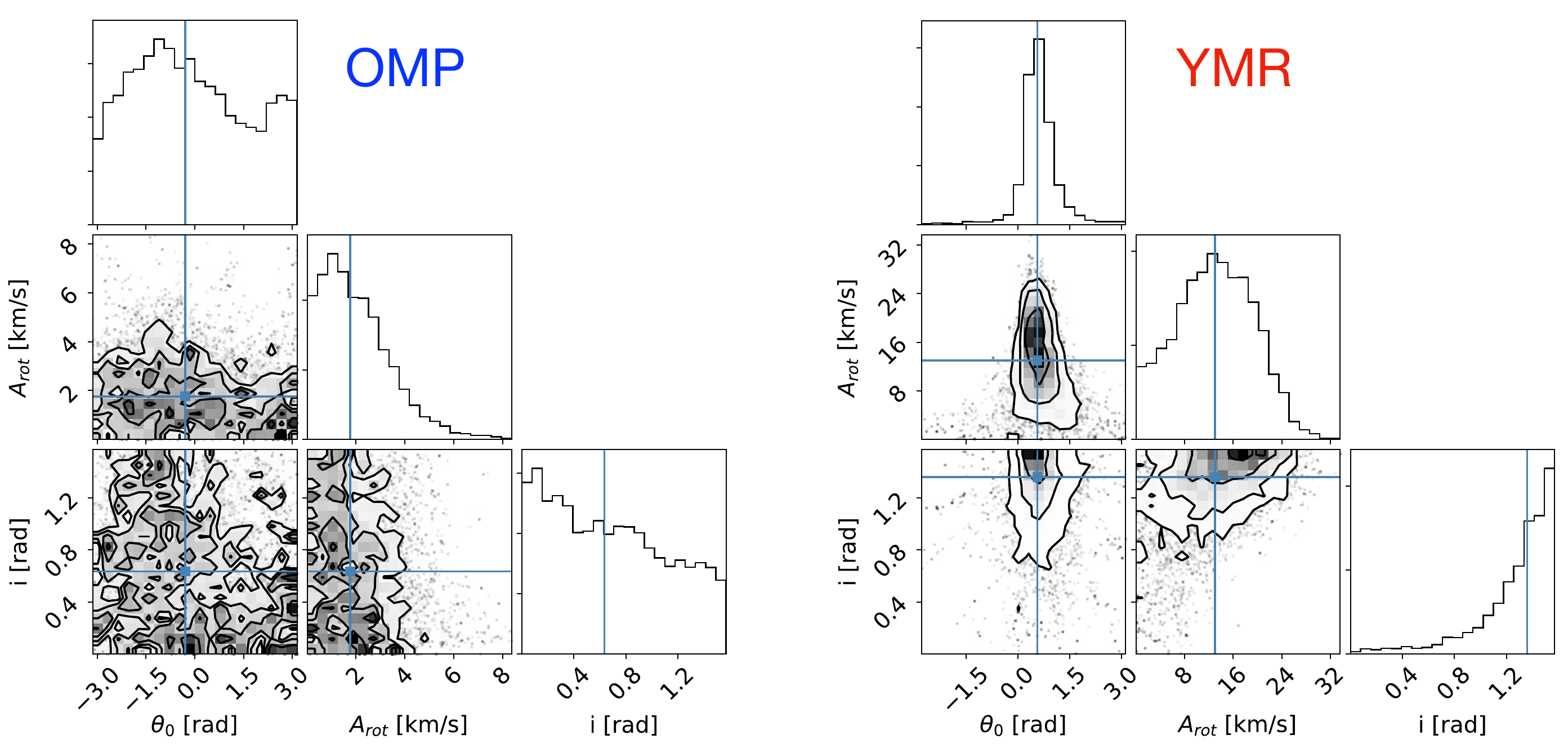}
\caption{Results of the MCMC sampling of intrinsic 3D rotation of the OMP component (\textit{left panel}; 108 data points) and of the YMR component (\textit{right panel}; 15 data points). The sampled  parameters are the position angle of the rotation axis in the plane of the sky ($\theta_0$), the amplitude of the rotation ($A_{rot}$), and the inclination angle of the rotation axis with respect to the line-of-sight ($i$). No rotation is observed for the OMP stars, while clear rotation, mostly along the line-of-sight velocity component, is measured for the YMR stars. The blue lines show the mean parameter value of the respective axis.}
\label{gaia2}
\end{figure*}

\begin{figure}
\centering
\includegraphics[scale=0.45]{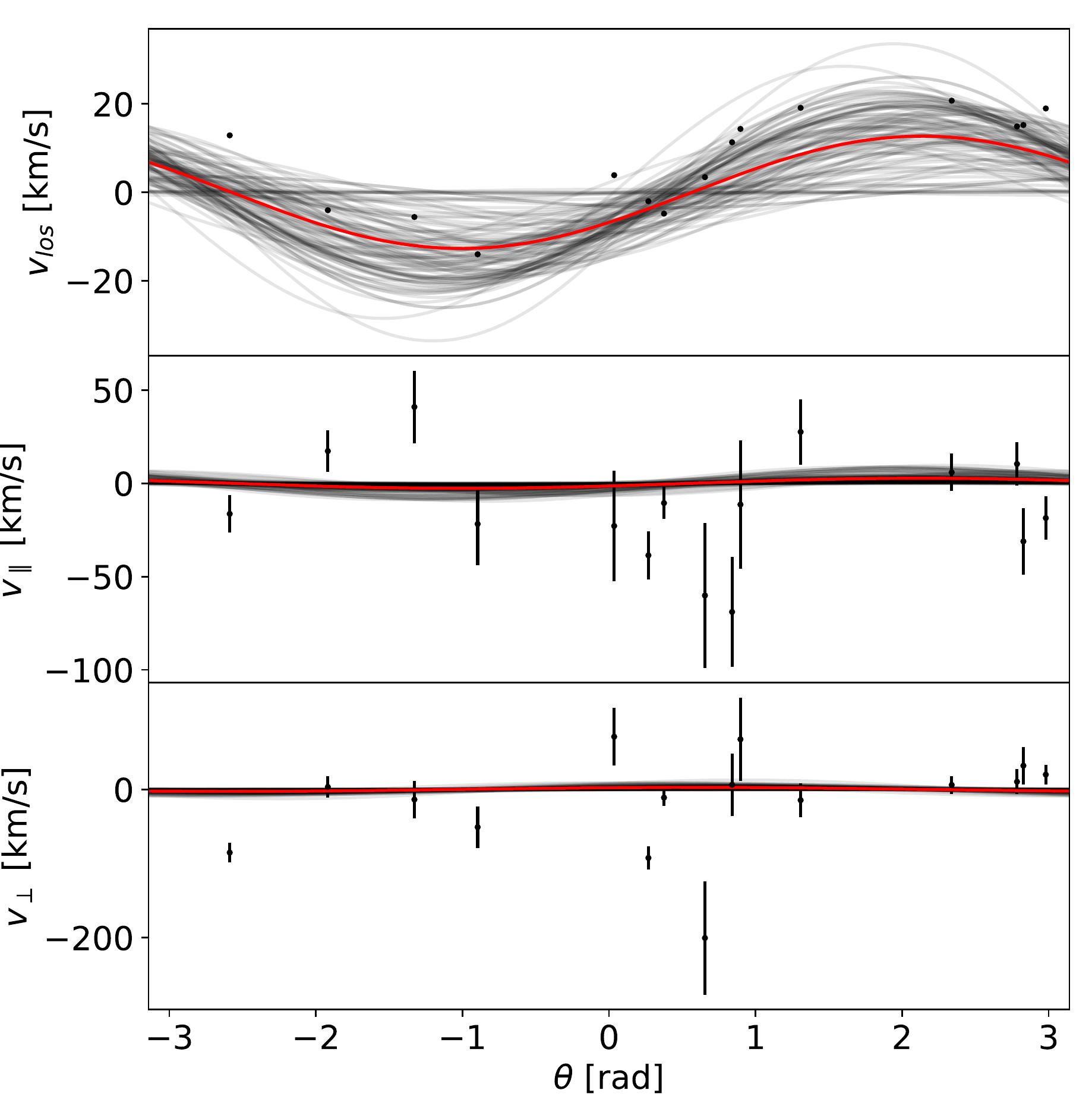}
\caption{Result of the fit for the YMR stars shown for the three components of the velocity vector, namely the line-of-sight component, the proper motion component parallel to the rotation axis, and the proper motion component perpendicular to the rotation axis. The red line indicates the result of the MCMC and the grey lines are 100 samples from the chain.}
\label{gaia3}
\end{figure}

The results obtained for the OMP and YMR components are shown in Figure\,\ref{gaia2} and reported in Table\,\ref{table_kin2}. We do not detect intrinsic rotation in the OMP component ($2.0^{+1.4}_{-1.8}$~km~s$^{-1}$). On the other hand, the YMR population shows a strong signature of rotation ($13.5^{+6.4}_{-5.8}$~km~s$^{-1}$), predominantly along the LOS (note that this signal is dominated by low-number statistics, as reflected by the large statistical uncertainties), since the recovered inclination angle $i$ is consistent with an edge-on view. 
In Figure\,\ref{gaia3}, we show the result of the simultaneous modeling of the three velocity components for the YMR component.

\begin{deluxetable}{c|ccc}
\tablecaption{3D modeling of the intrinsic rotation using the three velocity components.  \label{table_kin2}}
\tablehead{
\colhead{Population} & \colhead{$\theta_0$ (deg)} &   \colhead{$A_{rot}$ (km~s$^{-1}$)} & \colhead{$i$ (deg)} }
\startdata
  YMR & $ 32.4^{+24.6}_{-18.5}$    & $13.0^{+6.9}_{-6.6}$ & $77.9^{+9.1}_{-18.9}$       \\ 
  OMP & $-17.0^{+96.9}_{-124.0}$   & $1.77^{+1.1}_{-1.7}$  & $36.7^{+22.9}_{-32.7}$       \\ 
  YMR$^a$ & $ 29.8^{+18.9}_{-20.1}$    & $13.5^{+6.4}_{-5.8}$ & $80.8^{+6.9}_{-14.3}$       \\ 
  OMP$^a$ & $-40.1^{+78.5}_{-134.1}$   & $2.0^{+1.4}_{-1.8}$  & $29.8^{+19.5}_{-33.2}$       \\\hline
\enddata
\tablecomments{~ $^a$: For comparison, we add the values obtained assuming a distance to M54 of $28.4$~kpc \citep{Siegel2011}.}
\end{deluxetable}

\section{Comparison with $N$-body simulations} \label{simulations}

To understand the origin of the morphological and kinematical structure of the system, we simulated a two-component M54-like cluster with a total mass equal to $1.4\times10^6~M_\odot$. The modeled cluster is characterized by a YMR/OMP mass ratio of $0.20$ (based on the number ratio of stars), as suggested by the observations. The OMP population is represented by a non-rotating, spherical \citet{K66} profile with a total mass of $1.17\times10^6~M_\odot$, $W_0=8.6$, and a core radius of $0.72$\,pc \citep[][2010 edition]{Harris1996}. The remaining mass of the cluster is included as a flattened and centrally-concentrated YMR population (50\% of the mass is enclosed within $3$\,pc from the center of the cluster), that formed in situ from gas captured by the gravitational potential of the OMP component. The gas kept its angular momentum and formed a rotating stellar component. This disk-like component has an ellipticity ranging from $\epsilon\approx 0.1$ in the central region to $\epsilon=0.5-0.6$ at radii larger than $6$\,pc. The YMR population initially rotates with a peak velocity of $12$\,km~s$^{-1}$, a value similar to the maximum rotational velocity of the YMR population observed in M54. 

The $N$-body system was modeled using a total number of $N=50\,000$ ($N_{OMP}=41\,786$ and $N_{YMR}=8\,214$) of single mass particles. The mass of each particle is $m_*\approx28~M_\odot$ and we adopted a softening length of $0.01$\,pc to smooth the close encounters between particles. 
Taking into account both of these approximations, the simulation time was rescaled to the evolutionary time of the system formed by the actual number of stars, assuming an overall average stellar mass of $0.94~M_\odot$ (see Section~\ref{energy_eq}), by using the ratio between the relaxation times of the real and simulated system as described in \citet{Mastrobuono-Battisti2013}.
The initial conditions were built using the NEMO toolkit \citep{Teuben1995}.
We evolved the system for $2$\,Gyr -- the age of the YMR population -- using a version of the direct $N$-body code phiGRAPE adapted to run on GPUs \citep{Harfst2007, Gaburov2009}. The coevolution of the two populations and, in particular, the relaxation of the initially-flattened YMR component, modify the final shape and kinematics of the whole cluster.

\begin{figure*}
\centering
\includegraphics[width=0.45\textwidth]{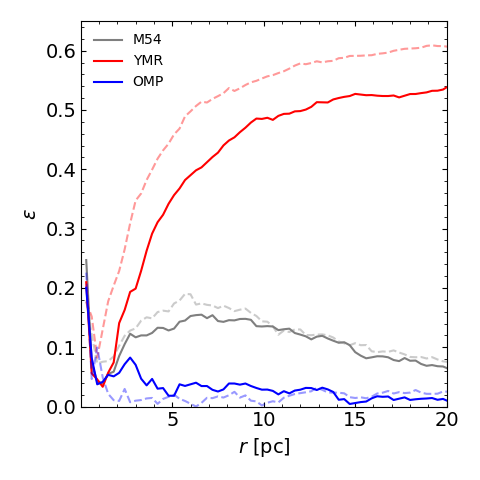}~\includegraphics[width=0.45\textwidth]{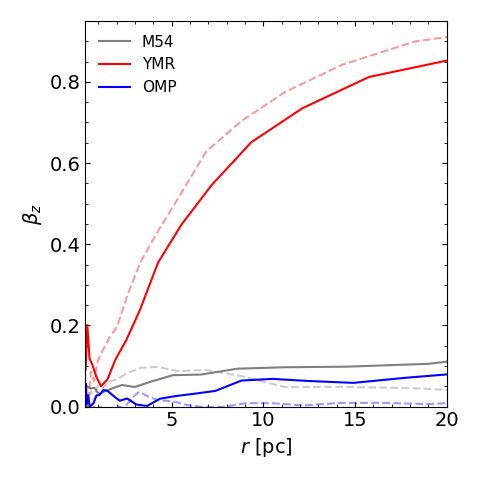}
\caption{Ellipticity (left panel) and anisotropy parameter $\beta_z$ (right panel) as a function of radius for the whole system (gray), the OMP (blue) and YMR (red) populations at 0~Gyr (dashed lines) and after 2\,Gyr (solid lines) of evolution as obtained from the $N$-body simulations. The angular momentum lost by the YMR population is acquired by the OMP population, that becomes slightly flattened and acquires a small amount of velocity anisotropy along the z axis. Plots are obtained considering the edge-on view of the cluster.}
\label{sim1}
\end{figure*}

To evaluate the ellipticity of the different populations at the end of the simulation, we calculated their axial ratios using the inertia moments, as detailed in \citet{Katz1991}. The $b/a$ ratio between the intermediate and major axis is approximately equal to $1.0$ at all radii, and the ellipticity is defined as $\epsilon=1-c/a$, where $c$ is the minor axis of the system. 
The final ellipticity of the YMR population varies significantly with the distance from the center of the cluster; after 2~Gyr of evolution, the YMR stars are still in a significantly flattened configuration with ellipticity close to the initial one. In particular, the young component is almost spherical in the central regions of the cluster, while the ellipticity increases up to $\epsilon=0.55$ at radii larger than 5\,pc. The ellipticity is $\approx 0.4$ at $7$\,pc, which is the half-mass radius of the cluster.  The slight decrease in the ellipticity of the young population corresponds to an increase of flattening in the old stellar component. 
After $2$\,Gyr of coevolution with the younger component, the initially-spherical OMP population becomes more flattened, reaching a maximum ellipticity of around $0.1$. The ellipticity of this population decreases steadily with the distance from the cluster center (see left panel of Figure \ref{sim1}). 

\begin{figure}
\centering
\includegraphics[width=0.45\textwidth]{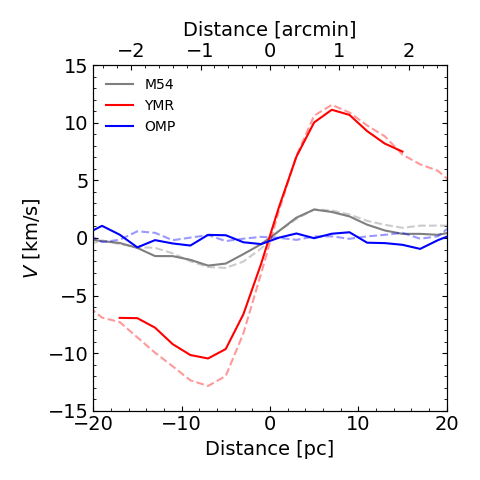}
\caption{Rotation curves for the whole system (gray), the OMP (blue) and YMR (red) populations at 0~Gyr (dashed lines) and after 2\,Gyr (solid lines) of evolution as obtained from the $N$-body simulations. While the YMR population still rotates significantly, the OMP population has acquired a small ($\sim1\,$km~s$^{-1}$) rotational speed. The cluster is seen edge-on.}
\label{sim2}
\end{figure}

\begin{figure*}
\centering
\includegraphics[width=0.45\textwidth]{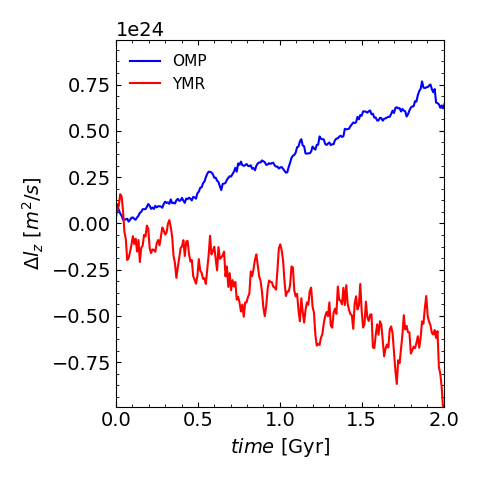}
\includegraphics[width=0.45\textwidth]{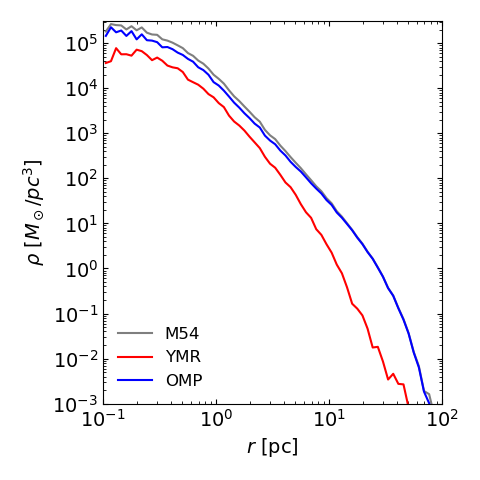}
\caption{Left: Evolution of the average angular momentum per unit mass perpendicular to the maximum rotation plane for the YMR (red) and OMP (blue) populations. The angular momentum lost by the YMR stars is redistributed among the OMP stars that, consequently, acquire a coherent rotational pattern and settle on a slightly flattened configuration.
Right: Spatial density profiles of the system, from the $N$-body simulations, considered as a whole (gray) and of the YMR (red) and OMP (blue) components taken separately as obtained from the $N$-body simulation.}
\label{sim3}
\end{figure*}

The change in morphology corresponds to a slight increase in the velocity anisotropy, whose amount is parametrized by the quantity $\beta_z=1-(\sigma_z/\sigma_R)^2$ in cylindrical coordinates with the $z$ axis parallel to the angular momentum vector of the system (see right panel of Figure \ref{sim1}). Thus, $\beta_z$ is related to how the system is flattened (in this case the system is oblate, given the positive value of $\beta_z$), depending on the ratio between the dispersion in the $z$ direction and in the plane perpendicular to the total angular momentum of the system.
While, after $2$\,Gyr, the YMR disc still rotates with a peak velocity of around $11$~km~s$^{-1}$, the OMP population only shows a weak rotation pattern, with a peak velocity of $\sim1\,$km~s$^{-1}$ (see Figure \ref{sim2}).  As already found by \citet{Mastrobuono-Battisti2013,Mastrobuono-Battisti2016}, this result suggests that the coevolution of an initially-spherical system with an embedded disc leads to an angular momentum redistribution between the flattened YMR and the OMP population. From this process follows the mixing of the YMR and the OMP populations and the increased flattening of the OMP population  (see left panel of Figure \ref{sim3} for the angular momentum evolution of the two populations). The growing flattening is accompanied by only a slight increase in the rotation speed of the OMP population. Finally, the two populations, after $2$\,Gyr of evolution, are not yet fully mixed, with the YMR population still more centrally concentrated and less extended than the OMP population (see right panel of Figure \ref{sim3}).

\section{Discussion} \label{discussion}

The kinematic results we present in this paper coupled to the dynamical simulations, add substantial evidence to the scenario we propose in Paper~I from the findings based on the stellar characterization analysis on M54 and its populations.

\subsection{Kinematic comparison}

As we mention in Paper~I and show in detail in this work, all populations are spatially coincident, as evidenced by their measured spatial and LOS velocity distributions \citep[see also][]{Dacosta1995, Ibata1997, Monaco2005b, Bellazzini2008}.
We observe that the three populations display different velocity dispersion profiles. For the OMP population the observed velocity dispersion is well described by a Plummer profile, with a median central velocity dispersion of $\sigma_0=15.30\pm0.54$~km~s$^{-1}$ and $\sigma\sim10$~km~s$^{-1}$ at $r=1\farcm6$. 
The profiles are close to flat for the YMR and IMR populations.
These are consistent with the velocity dispersion estimates from previous studies based on hundreds of stars \citep[e.g.][]{Bellazzini2008, Baumgardt2018}. 
The flat velocity dispersion profile observed for the YMR population could be explained by (i) energy equipartition (see Section~\ref{energy_eq}); (ii) the formation of the YMR population in a dynamically-cold component, which is consistent with being rotationally supported and with low-velocity dispersion; (iii) crowding effects due to the highly-crowded central region of M54 and the dominance of the OMP population over the YMR at these observed wavelengths. These effects limit our ability to extract an accurate number of stars that can produce an apparent drop in the central velocity dispersion.
We are working on chemo-dynamical models on M54 that will help us to better constrain these kinematic differences between populations considering, e.g., their different spatial distribution and amount of rotation.

The IMR population displays a central velocity dispersion consistent with the OMP population, however,  the profile is rather flat at a value of $\sim15$~km~s$^{-1}$.
This value at the largest radii is slightly higher than that found between $2\arcmin<r<8\arcmin$ by \citet{Bellazzini2008}, who showed the Sgr dSph stars followed a flat velocity dispersion profile at $\sim10$~km~s$^{-1}$. This difference may be due to our particular definition of the IMR population, which excludes younger stars (at preferentially low dispersion at large radii) which may be in the \citet{Bellazzini2008} sample. At larger radii until $\sim60\arcmin$, the Sgr dSph velocity dispersion profile \citep[see][]{Majewski2013} seems to be consistent with other dSph galaxies around the Milky Way \citep[e.g.,][]{Walker2007}. 
The flat dispersion profile of the IMR and in the outskirts of Sgr dSph, and its similarity to other dSphs could have several origins. For example in classical dSphs this is interpreted as a result of large dark matter dominated potentials, alternatively it could be due to tidally stripped or potential escaper stars \citep{Bianchini2019}. In such a complex system, the details of tidal disruption, including how much baryonic and dark matter remain in the various volumes and the joint effect on velocity dispersion profiles, are far from certain for this specific case. A large radial coverage of Sgr dSph along with detailed simulations is likely needed to understand these processes. Unfortunately, we cannot constrain these effects since our data is confined to the innermost $2\arcmin$, corresponding to the nuclear region which might or might not be affected by tidal effects yet.

Our large MUSE data set allowed us to extract precise kinematics by using a large number of stars for each population. We find that the YMR population rotates ($A_{rot}=4.91\pm0.90$~km~s$^{-1}$ and $V_{max}=7.98\pm1.01$~km~s$^{-1}$) at a considerably higher speed than the OMP, which shows a weak sign of rotation ($A_{rot}=0.75\pm0.55$~km~s$^{-1}$ and $V_{max}=1.58\pm1.14$~km~s$^{-1}$). 
Note that the rotation obtained from 3D kinematics (see Table~\ref{table_kin2}) is higher than the one obtained with the LOS data only (yet, consistent within the uncertainties). This difference might be the result of the subsample of stars from the MUSE sample for the estimate using the \textit{Gaia} data where we obtain large error bars. Additionally, the 3D case does not find exactly an edge-on angle ($\sim80\degree$), which can increase the intrinsic $A_{rot}$.

We note that \textit{Gaia} proper motions are affected by systematic errors, which have the same order of magnitude as the rotation signal we are trying to measure ($0.07$~mas/yr, $8-9$~km~s$^{-1}$ at M54's distance). 
For M54 the uncertainties of \textit{Gaia} proper motions are fairly large, due to the rather large distance and the high stellar density at the center. This reduces the accuracy of proper motions measurements, which are typically found to have errors of the order of $\sim$20\,km~s$^{-1}$. All of these factors can effectively hide any rotation signal present in the plane of the sky. Future \textit{Gaia} data releases are needed to obtain more accurate 3-D rotation results. There are HST proper motions for M54 \citep{Bellini2014,Watkins2015a}, however, they are relative, not absolute measurements, and cannot be used to study rotation.
 
Using \textit{Gaia DR2} data of stars in both populations, we found that the OMP and YMR populations are comoving in 3-D space. This suggests these two populations are dynamically bound, adding substantial evidence for spatial coincidence and against chance alignment.

\subsection{YMR: evidence of in situ formation} \label{YMR}

Young sub-components in NSCs have been detected in different types of galaxies (e.g., early- and late-type). These young structures are more centrally concentrated and flattened \citep{Seth2006, Carson2015, Feldmeier-Krause2015, Nguyen2017}. A similar central concentration was found in $\omega$\,Cen \citep{vandeVen2006}, a presumed former NSC.

In Paper~I, we found evidence to suggest that the YMR population formed $\sim$3\,Gyr ago likely in situ in a star-forming gas disk. In addition, we found that this population is the most centrally concentrated and the most flattened \mbox{($\epsilon=0.31\pm0.10$)}.

In this paper, we add that the YMR population has a rotation amplitude of $\sim$5\,km~s$^{-1}$, reaching a maximum rotation of $\sim8$~km~s$^{-1}$, consistent with the high degree of flattening that we found. In contrast, the OMP population shows a small rotation of 0.7\,km~s$^{-1}$, but it shows an amount of flattening similar to other clusters with higher rotation ($\epsilon\sim0.15$). We will discuss this in the next section. From $N$-body simulations, \citet{Mastrobuono-Battisti2013, Mastrobuono-Battisti2016,Mastrobuono-Battisti2019b} found that the young structures can survive despite being embedded in an older population. Moreover, they are found to rotate faster than the population in which they are embedded. Given the significantly different kinematics of these populations it appears likely that the small difference observed in ellipticity is real. A representative case of a young stellar disk is observed in the Milky Way NSC \citep{Genzel2003, Paumard2006, Bartko2009, Lu2009, Lu2013, Yelda2014, Feldmeier-Krause2015}.

\subsection{Kinematic effects on the OMP stars by the YMR} \label{omp+ymr}

The formation and kinematics of the YMR population can affect the kinematics and shape of the population in which they were born.
We found that the OMP population has a low rotation of $A_{rot}=0.75\pm0.55$\,km~s$^{-1}$, with an ellipticity of  $\epsilon=0.16\pm0.06$. In the ($V/\sigma_0,\epsilon$) diagram (see Figure~\ref{diagnostic}) we observe that the OMP population (blue square) is more flattened than most of the GCs rotating at a similar speed, and is similar to the most flattened clusters, e.g., $\omega$\,Cen (black diamond), M22 (black pentagon), which seem to rotate even faster.

Our $N$-body simulations consider a disk-like centrally-concentrated YMR population, embedded in an initially-spherical and five times more massive OMP population. The two populations co-evolve for $2$~Gyr, which is the estimated age of the young population observed in M54. At the end of the simulation, the YMR population, which is born in a disc because of the angular momentum of the progenitor gas, relaxes and evolves towards a more spherical configuration while redistributing its angular momentum among the OMP stars. As a consequence of this process, the YMR disc slows down its rotation. At the same time, the OMP population acquires the angular momentum lost by the younger stars, decreases its velocity dispersion in the $z$ direction and becomes slightly more flattened. This phenomenon could, at least partially, explain the ellipticity observed for the OMP population. Previous stellar populations, born in a disc, could have also contributed to the flattening of the OMP population leading to a more significant final effect, and shaping the present morphology of the OMP population. The relative angular momentum and ellipticity evolution due to the YMR interacting with the OMP population might be even stronger than our simulations suggest, given that in Section~\ref{energy_eq}, the average mass of stars in the YMR population is approximately two times more than that in the OMP population, compared to the 1:1 ratio adopted in our simulations. Further simulations will help to address this effects.

Past mergers could have increased the ellipticity of the system, but would have also significantly affected its kinematics, producing a higher rotational signal than that observed in M54 \citep{Tsatsi2017, Mastrobuono-Battisti2019}. If the YMR population formed in a central disc that relaxed, leading to the observed OMP population flattening, we predict a radially-increasing velocity anisotropy for the YMR population, with $\beta_z$ down to values smaller than $0.6$ inside the half-mass radius of the cluster (see Figure~\ref{sim1}). The OMP population has a slight radial increase in the anisotropy profile. Future \textit{Gaia} observations will be able to verify these predictions at least in the outskirts of the cluster.

In Paper~I, we found different spatial distributions for the OMP and YMR populations, with the YMR being the most centrally concentrated.  \citet{Mastrobuono-Battisti2013, Mastrobuono-Battisti2016} also found that even after 12\,Gyr the two structures are not yet fully mixed, i.e. are still showing different distributions. In the simulations presented in this work, we also observe that after 2\,Gyr of evolution (close to the age of the YMR population) the two populations are not yet fully mixed, and the YMR population is found to be still more centrally concentrated than the OMP population.

Note that the $N$-body simulations are not meant to exactly reproduce the current properties of M54, but rather to test the possibility that the interactions between the OMP and YMR populations might have mutually modified the dynamical configuration of these two populations. 

\subsection{OMP: remnant of a cluster merger?} \label{omp}

In Paper~I, we found that the OMP population displays a large spread in both metallicity and age. A large spread in metallicity alone can be explained by self-enrichment during formation \citep{Bailin2018}, but it does not explain the large spread in age. Thus, we suggested that this could be the result of a merger event between two or more clusters that fell into the central region of the host, as suggested through simulations \citep[e.g.,][]{Amaro-Seoane2013,Gavagnin2016,Bekki2016,Khoperskov2018,Mastrobuono-Battisti2019}. 
In addition, a contribution to this high spread in metallicity might be due to the old, metal-poor population of Sgr dSph that mixed with the merged GCs to the resulting OMP population of this NSC.

According to the $N$-body simulation presented in this work, the formation and evolution of the YMR population alone can explain the shape and kinematics of the OMP. However, it does not explain the large spread in both age and metallicity, that we found in this population.

The final rotation of two merged clusters strongly depends on the conditions, as the orbital configuration and relaxation states of the merging clusters do not always result in a highly rotating structure \citep{Mastrobuono-Battisti2019}. Based on the kinematics, we cannot be certain that two merging clusters were actually involved in the formation of the OMP population but cannot discard this possibility either. 
Further high-precision observations and more detailed simulations appropriate to mergers in the Sgr potential may help to shed light on whether this process may have contributed to the build-up of the OMP population in the Sgr dSph.

\subsection{Comparison with $\omega$\,Cen}

The similarities between NSCs in dwarf galaxies and high-mass, metal complex GCs suggest that such GCs might be former nuclei of dwarf galaxies accreted by the Milky Way \citep{Zinnecker1988,Boker2008,daCosta2016}. One of the most likely former nuclei is $\omega$~Cen, which is the most massive GC in the Milky Way \citep[$3.55\pm0.03\times10^6~M_\odot$, ][]{Baumgardt2018}. It has a large spread in both iron \citep{Johnson2010} and age \citep{Hilker2004, Villanova2014} among its stars and hosts multiple stellar populations \citep{Milone2017, Bellini2017}, which present different kinematics \citep{Bellini2018}.
These properties, and their similarity to what we have found in M54, make a strong case for $\omega$~Cen to be a stripped NSC that once resided in a dwarf galaxy now accreted by the Milky Way \citep{Lee1999,Majewski2000,Bekki2003,Carretta2010a,Majewski2012}.
Using $N$-body simulations, \citet{Ibata2019_nature} found that the ``Fimbulthul" structure detected with the \textit{Gaia DR2} observations \citep{Ibata2019} could be a tidal stream of $\omega$~Cen. This suggests that there may be stars in the halo that were once part of $\omega$~Cen; finding these escaped stars would help us to understand the progenitor. \citet{Massari2019} used kinematic information of the Milky Way GCs from \textit{Gaia} to suggest that $\omega$~Cen could be the former NSC of the progenitor galaxy of the merger event \textit{Gaia}-Enceladus \citep{Helmi2018}.
Observations of $\omega$~Cen and M54 suggest that $\omega$~Cen could be in a more advanced ``stripped nuclei stage" \citep[see][]{Pfeffer2013} than M54 \citep[e.g.,][]{Bellazzini2008,Carretta2010a}, since M54 is still observed at the photometric center of its host galaxy \citep{Ibata1994}, which is in ongoing disruption by the tidal field of the Milky Way \citep{Ibata1997}.

Similar to other GCs, $\omega$~Cen presents ellipticity variations as a function of radius \citep{Geyer1983}, and is found to be one of the most flattened Milky Way GCs \citep{Meylan1987}. As we mentioned in Section~\ref{rotation_vs_ellipticity}, there is a correlation between the amount of rotation of a cluster and its ellipticity \citep{Fabricius2014, Kamann2018}, which has been detected for $\omega$~Cen \citep[e.g.,][]{Meylan1986,Bianchini2013,Bianchini2018,Kamann2018}. 
$\omega$~Cen behaves close to an isotropic oblate rotator in the inner parts ($<10\arcmin$), becoming more anisotropic at larger radii, probably due to the tidal effects by the Milky Way \citep{vandeVen2006} which do not affect the inner parts of the cluster.
For M54, we find a difference of $\Delta\epsilon\approx0.05$ between the ellipticity estimated over the entire FOV ($\sim2.5\times$ R$_{\mathrm{HL}}$) and at the half-light radius. Being cautious with the ellipticity uncertainties, this suggests that the ellipticity for M54 varies as a function of radius, becoming more flattened at larger radii.
From our (V$/\sigma_0,\epsilon$) diagram (see Figure~\ref{diagnostic}) we observe that M54 is close to the isotropic oblate rotator relation when looking at its half-light radius, showing low rotation with a small degree of flattening. 
\citet{Watkins2015a} used proper motions to show that M54 is isotropic out to the half-light radius. However, this does not imply that the cluster cannot display anisotropy at larger radii.
For the observations of M54, further analysis and information are needed to confirm and constrain its presumed radial anisotropy.

Additionally, a disk-like component was detected in $\omega$~Cen by \citet{vandeVen2006}. We found a similar structure in M54 corresponding to the YMR population that shows a strong rotational signal, and that most likely formed in the center of the OMP population. We speculate that also the disk-like component in $\omega$~Cen was born within the old population, which can be tested by measuring its metallicity, age and rotational structure.

Although small differences exist between $\omega$~Cen and M54, the similarity of their stellar populations, their morphology and kinematics provides additional evidence that $\omega$~Cen is, in fact, the nucleus of a stripped dwarf galaxy. This comparison considers M54 as the NSC of Sgr dSph, which is now composed of three main populations. This nuclear stellar structure is what we would observe if its host galaxy, the Sgr dSph, was totally stripped.

\section{Conclusions} \label{conclu}

In Paper I, we identified three subpopulations in M54, the NSC of the Sgr dSph based only on their ages and metallicities. They are: (i) YMR: young metal rich, $2.2$\,Gyr and \mbox{[Fe/H]$=-0.04$}, (ii) IMR: intermediate-age metal rich, $4.3$\,Gyr and \mbox{[Fe/H]$=-0.29$}, and (iii) OMP: old metal poor, $12.2$\,Gyr and \mbox{[Fe/H]$=-1.41$}. Here we have presented a kinematic analysis of each subpopulation to both characterize the populations and gain new insight into their formation and evolution.

In this paper, we continued the characterization of the stellar populations in M54 by adding kinematic information. M54 is one of the closest extragalactic NSC, at a distance where stars can be resolved and characterized, enabling the distinction of the stellar populations. In studies of extragalactic NSCs, the different stellar populations cannot be easily resolved, restricting the kinematic analysis to the nucleus as one single structure.
In M54 we have the opportunity to directly compare the kinematics for the nucleus considering all the stars in the sample (i.e. like one single structure), as is usually done for extragalactic NSCs, and for the different populations. We present the kinematics for two cases: (i) all M54 star members with no distinction, and (ii) for the populations detected in M54 in Paper~I.

We found that all three populations show different velocity dispersion profiles; the YMR and IMR populations follow a flat distribution, whereas, the OMP population is well-defined by a Plummer profile. We find that all populations show a different amount of rotation. The YMR population shows a considerable amount of rotation ($\sim5$~km~s$^{-1}$), followed by the IMR population, which also rotates but more slowly ($\sim2$~km~s$^{-1}$). We detect a weak signal of rotation for the OMP population ($<1$~km~s$^{-1}$). 
Hence, the findings from the kinematic analysis support our findings from Paper I that the populations do not have a common origin. 

From these findings together with those in Paper~I, we suggest the following conclusions:

\begin{itemize}

\item From our large sample of stars, we find that all populations have the same systemic velocity. Combining our data with proper motions from \textit{Gaia DR2}, we find that the stars in the OMP and YMR population are comoving in 3D space. Taken together, this secures the finding that the populations are spatially coincident and disfavours chance alignment.

\item The YMR population displays a high amount of rotation. Combined with the fact that it is more flattened and more centrally concentrated, this strongly favors the scenario wherein these stars formed in-situ from enriched gas accreted at the center of M54. 

\item The OMP population appears more flattened than typical GCs rotating at the same speed.
The $N$-body simulation emulating the YMR-OMP system in M54 suggests that this could result from the angular momentum transferred from the YMR to the OMP population, thus decreasing its velocity dispersion in the $z$ direction and becoming more flattened. 
However, this alone does not explain the high iron and age spread this population presents, which we suggested to be signs of a merger event between two or more GCs in the center of the host. 
The merger scenario cannot be ruled out since mergers of GCs both do or do not end in a rotating structure depending on the initial conditions. 
The OMP population might also be the result of a GC that sank to the center where the old, metal-poor population from Sgr dSph was already in place, thus together contributing to the high spread in age and metallicity.

\item We tested the possibility that effects connected to energy equipartition could imprint different kinematics due to energy exchange between stars with different masses. We find that the difference in velocity dispersions between the YMR and OMP populations are consistent with equipartition effects driven by relaxation processes.

\item The current information we obtain from the IMR population does not allow us to constrain if its stars are dynamically bound to the nucleus. However, the kinematics of this population are consistent with these stars being part of the central field of the Sgr dSph.

\end{itemize}

The kinematic analysis presented in this paper certainly adds an essential piece of information to the understanding of M54 and its populations. 

NSCs display multiple populations - as observed in the nucleus of the Milky Way - and grow with time depending on their environmental conditions.
The nucleus of the Sgr dSph possibly started with an old and metal-poor population where one or more GCs merged, thus forming together an old, metal-poor stellar structure with a considerable spread in age and metallicity. In spite of our limited ability to differentiate them, this is what we observe today as the OMP component of the nucleus of the Sgr dSph. Then, $\sim3$~Gyr ago a young population (YMR) was born within the nucleus. All these populations in this NSC that Messier observed for the first time in 1778 and named M54 form the NSC in the Sgr dSph. This structure with all its populations is likely what we would observe after the total stripping of the host, as in the case of $\omega$~Cen.

\section{Future prospects} \label{future}

The confirmation of the putative IMBH at the centre of M54 would provide a strong piece of evidence to support its nuclear origin. It would also imply their existence in other NSCs in low-mass galaxies where they have not been detected, possibly due to instrumentation limitations.
Many high-mass metal complex GCs are IMBH host candidates, but there is no confirmation yet, e.g., $\omega$~Cen \citep{Noyola2008, Jalali2012, Baumgardt2017}, G1 \citep{Gebhardt2005}, NGC~6388 \citep{Luetzgendorf2011}.  M54 is presumed to host an IMBH of $10^4~M_\odot$  \citep{Ibata2009, Baumgardt2017}, but a confirmation is still needed.

We are performing dynamical modeling of M54 and its individual stellar populations to constrain the existence and mass of the IMBH in its center. This will also constrain variations in anisotropy as a function of radius. To achieve this, we work with an additional MUSE narrow-field mode data set which allows us to extract hundreds of stars in the innermost $3\arcsec$ of M54, in the sphere of influence of the presumed IMBH. In an upcoming paper, we will present discrete Jeans dynamical models for M54 where we expect to better constrain its dynamical properties and the existence of the IMBH.

\acknowledgments{
SK gratefully acknowledges funding from a European Research Council consolidator grant (ERC-CoG-646928- Multi-Pop). RL acknowledges funding from a Natural Sciences and Engineering Research Council of Canada PDF award and AMB, RL and NN acknowledge support by Sonderforschungsbereich SFB 881 ``The MilkyWay System" (subproject A7, A8 and B8) -- Project-ID 138713538 -- of the German Research Foundation (DFG).  MAC, RL and PB acknowledge support from DAAD PPP project number 57316058 ``Finding and exploiting accreted star clusters in the Milky Way". Work by A.C.S. on this project was supported by NSF grant AST-1350389. GvdV and LLW acknowledge funding from the European Research Council (ERC) under the European Union's Horizon 2020 research and innovation programme under grant agreement No 724857 (Consolidator Grant ArcheoDyn). This research has made use of NASA's Astrophysics Data System. This work has made use of data from the European Space Agency (ESA) mission {\it Gaia} (\url{https://www.cosmos.esa.int/gaia}), processed by the {\it Gaia} Data Processing and Analysis Consortium (DPAC,
\url{https://www.cosmos.esa.int/web/gaia/dpac/consortium}). Funding for the DPAC has been provided by national institutions, in particular the institutions participating in the {\it Gaia} Multilateral Agreement.
}

\bibliographystyle{yahapj}
\bibliography{mybib}{}

\end{document}